\documentclass[a4paper,11pt]{article}
\pdfoutput=1 % if your are submitting a pdflatex (i.e. if you have
\usepackage{jcappub} % for details on the use of the package, please
\usepackage{subcaption}
                     
\usepackage{svg}
\newcommand{\orcid}[1]{\href{https://orcid.org/#1}{\includesvg[width=10pt]{orcid}}}
\newcommand{\github}[1]{\href{https://github.com/#1}{\includesvg[width=10pt]{github}}}

\usepackage{booktabs}

\usepackage[capitalise]{cleveref}
\usepackage{braket}

\usepackage[T1]{fontenc} % if needed

\usepackage{slashed}
\usepackage{url}
\usepackage{soul}

\title{\boldmath Long Range Interactions in Cosmology: Implications for Neutrinos}
\author[a,b,c]{Ivan Esteban \orcid{0000-0001-5265-2404},}
\author[c]{Jordi Salvado \orcid{0000-0002-7847-2142}}

\hypersetup{pdfinfo={
Author={I.~Esteban, J.~Salvado},
Keywords={Beyond the Standard Model, Long Range Interactions, Cosmology, Neutrinos, Neutrino Interactions, Neutrino Masses}
}}

\affiliation[a]{Center for Cosmology and AstroParticle Physics (CCAPP), Ohio State University, Columbus, OH 43210}
\affiliation[b]{Department of Physics, Ohio State University, Columbus, OH 43210}
\affiliation[c]{Departament de Física Quàntica i Astrofísica and
  Institut de Ciències del Cosmos, Universitat de Barcelona, Diagonal
  647, E-08028 Barcelona, Spain}

\emailAdd{esteban.6@osu.edu}
\emailAdd{jsalvado@icc.ub.edu}

\abstract{Cosmology is well suited to study the effects of long range
  interactions due to the large densities in the early Universe. In
  this article, we explore how the energy density and equation of
  state of a fermion system diverge from the commonly assumed ideal
  gas form under the presence of scalar long range interactions with a
  range much smaller than cosmological scales. In this scenario,
  ``small''-scale physics can impact our largest-scale
  observations. As a benchmark, we apply the formalism to
  self-interacting neutrinos, performing an analysis to present and
  future cosmological data. Our results show that the current
  cosmological neutrino mass bound is fully avoided in the presence of
  a long range interaction, opening the possibility for a
  laboratory neutrino mass detection in the near future. We also
  demonstrate an interesting complementarity between neutrino
  laboratory experiments and the future EUCLID survey.   
}
\begin{document}
\maketitle
\flushbottom

\section{Introduction}

The quest for exploring new fundamental interactions has traditionally focused on high-energy probes such as 
particle colliders. The main hypothesis underlying these searches is that new physics has evaded detection because it only acts 
at short distances or, in particle physics terms, is mediated by heavy states. However, new physics could be 
mediated by light particles (i.e., it could have a long range) and remain undiscovered simply because it couples 
too weakly to matter. 

In this case, small couplings can be overcome by setups with large amounts of particles over which effects 
accumulate coherently~\cite{Fischbach:1996eq,Adelberger:2003zx,Williams:2004qba,Adelberger:2009zz,Wagner:2012ui}. An archetypal example is gravity: despite being exceptionally weak, it was the first 
fundamental force discovered as it adds up over all particles in macroscopic objects. Cosmology is particularly 
well suited to explore this sort of many-particle effects, as in the early Universe particle number densities were extraordinarily high --- at 
Big Bang Nucleosynthesis, for instance, as large as ${\mathcal{O}(\mathrm{MeV}^3) \sim 10^{32} \, 
\mathrm{cm}^{-3}}$. Furthermore, cosmological evolution is dominated
by gravity, and it is thus 
susceptible to be modified by any stronger interaction.

Along this line, long range interactions would directly impact cosmological structure formation, a hypothesis 
that has been widely explored in modified gravity and fifth force
scenarios~\cite{Friedman:1991dj,Bean:2001ys,Gubser:2004uh,Nusser:2004qu,Bean:2008ac,Kesden:2006zb,Bai:2015vca,Mohapi:2015gua,Zhang:2007nk,Amendola:2007rr,Avilez:2013dxa,Li:2013nwa,Umilta:2015cta,Acquaviva:2007mm,Alonso:2016suf,Sola:2019jek,Sola:2020lba}. But cosmology is also sensitive to the total energy density and pressure of the Universe. These are commonly computed 
assuming that the homogeneous and isotropic cosmological fluid behaves as an ideal gas. Nevertheless, this 
assumption breaks down under the presence of an interaction whose range is larger than the interparticle distance. In 
this work, we will consistently study such effects and their observable consequences in cosmology. We will focus 
on interactions among fermions mediated by scalar fields, as they are simple and universally attractive. That is, their 
effects accumulate over all particles and antiparticles with any spin orientation.

On top of that, we will face these models with the current precise cosmological data. To this purpose, we 
will focus on long range interactions among neutrinos. These particles
are abundantly produced in the early Universe, significantly affecting
its evolution, but their self interactions are poorly constrained~\cite{Lessa:2007up,Pasquini:2015fjv,Blinov:2019gcj,Agostini:2015nwa,Blum:2018ljv,Brune:2018sab,Brdar:2020nbj}. 
The cosmological impact of putative neutrino self interactions has been widely studied~\cite{Beacom:2004yd,Chacko:2003dt,Chacko:2004cz,Hannestad:2004qu,Hannestad:2005ex,Bell:2005dr,Friedland:2007vv}, and lately there has 
been a renewed interest due to their possible relationship to the Hubble tension~\cite{Archidiacono:2016kkh,DiValentino:2017oaw,Kreisch:2019yzn,Park:2019ibn,Forastieri:2019cuf,Escudero:2019gvw,Ghosh:2019tab,Mazumdar:2020ibx,Choudhury:2020tka,Brinckmann:2020bcn} or the short 
baseline neutrino anomalies~\cite{Dasgupta:2013zpn,Hannestad:2013ana,Chu:2015ipa,Cherry:2016jol,Forastieri:2017oma,Song:2018zyl,Chu:2018gxk,Choudhury:2020tka,Brinckmann:2020bcn}. The interactions
explored in the literature are generically mediated by heavy
particles, but if the mediator is light long range effects need to be
taken into account. And, in fact,
electroweak gauge invariance along with the absence of new physics in
the charged lepton sector suggests that new neutrino interactions 
should be mediated by light particles~\cite{Blinov:2019gcj,Berbig:2020wve,Farzan:2015doa,Farzan:2015hkd}.

The study of neutrino properties is also an interesting topic by
itself, as the observation of mass-induced neutrino flavour transitions constitutes our first 
laboratory evidence for physics beyond the Standard
Model~\cite{Pontecorvo:1967fh,Gribov:1968kq,GonzalezGarcia:2007ib}. In particular, measuring the absolute neutrino mass 
scale is the holy grail of neutrino physics, as it would be a hint towards a new energy scale of Nature. 
Cosmology is particularly appropriate for this purpose, because massive neutrinos should become 
non-relativistic at times in which they impact Cosmic Microwave Background (CMB) and Large Scale Structure (LSS) data. Future surveys aim to pin down the absolute 
neutrino mass scale at the {$\gtrsim 3\sigma$} level~\cite{Laureijs:2011gra,Hamann:2012fe,Aghamousa:2016zmz,Sprenger:2018tdb}, much more precisely than 
current and near future laboratory experiments~\cite{Aker:2019uuj,Monreal:2009za,Gastaldo:2013wha,Alpert:2014lfa,Croce:2015kwa}. Relaxing the 
cosmological neutrino mass bound has been a subject of intensive research~\cite{GonzalezGarcia:2010un,Cuoco:2005qr,Oldengott:2019lke,Bellomo:2016xhl,Vagnozzi:2017ovm,Vagnozzi:2018jhn,Gariazzo:2018meg,Chacko:2019nej,Escudero:2020ped}, particularly because if taken at face value the 
current bound from CMB data~\cite{Aghanim:2018eyx} implies that the neutrino mass scale is
beyond the reach of present 
and near future laboratory experiments. As we will see, neutrino long range interactions drastically affect this bound.

This article is structured as follows. In \cref{sec:formalism} we introduce our formalism for a generic self-interacting 
fermion. We study and numerically solve the equations of motion for the homogeneous and isotropic case in 
\cref{sec:homogeneous}, and in \cref{sec:perturbations} we consider linear perturbations and their stability. In 
\cref{sec:analysis} we focus on neutrino long range interactions, showing their impact on current cosmological data 
(\cref{sec:data_now}) as well as the future prospects with the LSS EUCLID survey (\cref{sec:data_future}). We summarize our results and conclude in \cref{sec:conclusions}.

\section{Formalism}
\label{sec:formalism}

As discussed in the Introduction, we will study scalar-mediated long range interactions among fermions. The action of the 
system is given by\footnote{The formalism is similar to that of models
where dark energy is induced by neutrinos or dark matter~\cite{Fardon:2003eh,Kaplan:2004dq,Gu:2003er,Peccei:2004sz,Bean:2000zm,Bean:2001ys}. Nevertheless, in those models the scalar field
has a more exotic potential to mimic dark energy. 
Because of this, our scenario has different physical consequences.}
\begin{equation}
S = \int \sqrt{-\mathcal{G}} \, \mathrm{d}^4 x \left(- \frac{1}{2} D_\mu  \hat\phi D^\mu  \hat\phi - \frac{1}{2} M_\phi^2  \hat\phi^2 + i \bar{\psi} \slashed{D} \psi - m_0 \bar{\psi} \psi - g  \hat\phi \bar{\psi} \psi \right) \, ,
\label{eq:Lagrangian}
\end{equation}
where $\hat{\phi}$ and $\psi$ are the scalar and fermion fields
respectively, $\mathcal{G}$ is the determinant of the metric and 
$D_\mu$ its associated covariant derivative, $M_\phi$ and $m_0$ are
the scalar and fermion masses respectively, $g$ is the interaction
coupling, and we have used the metric signature $(-, +, +, +)$. The equations of motion immediately follow
\begin{align}
-D_\mu D^\mu  \hat\phi + M_\phi^2 \hat\phi & = - g \bar{\psi} \psi \, ,
\label{eq:scalarMasterEq} \\
i \slashed{D} \psi - (m_0 + g \hat\phi) \psi & = 0 \label{eq:diracEq}\, .
\end{align}

\Cref{eq:scalarMasterEq,eq:diracEq} are equations 
for the quantum fields $\hat\phi$ and $\psi$. However, as we are interested 
in studying cosmological scales, the coherence length of the fermion
field is generically much 
smaller than any considered distance. Thus, it can be analyzed in terms of a
phase space distribution $f(x^\mu, P_\mu)$ of classical particles with positions $x^\mu$ and
conjugate momenta $P_\mu$. Analogously, the sourced scalar field will
generically have a large 
occupation number, and can be well described by a classical field $\phi(x^\mu)$.

In this limit (see \cref{sec:app_classical} for the details), the Dirac-like 
equation~\eqref{eq:diracEq} gives the dispersion relation for the fermions
\begin{equation}
    P_\mu P^\mu = - \tilde{m}^2 \, ,
\end{equation}
where
\begin{equation}
    \tilde{m} \equiv m_0 + g \phi
\end{equation}
is the effective fermion mass. The classical limit of \cref{eq:scalarMasterEq} reads
(see \cref{sec:app_classical})
\begin{equation}
-D_\mu D^\mu  \phi + M_\phi^2 \phi = - g \int \frac{\mathrm{d}P_1 \mathrm{d}P_2 \mathrm{d}P_3}{\sqrt{-\mathcal{G}}} \frac{\tilde{m}}{P^0} f(x^\mu, P_\mu) \, ,
\label{eq:scalarKGexplicit_conjugate}
\end{equation}
where the right hand side corresponds to the expectation value of $\bar{\psi} \psi$ evaluated for the 
fermion state. If the space-time components of the metric vanish,
we can rewrite \cref{eq:scalarKGexplicit_conjugate} in terms of the physical momentum $\vec{p}$ as
\begin{equation}
-D_\mu D^\mu \phi + M_\phi^2 \phi = - g \int \mathrm{d}^3 \vec{p} \frac{m_0 + g \phi}{\sqrt{|\vec{p}|^2 + (m_0 + g \phi)^2}} f(x^\mu, P_\mu(\vec{p})) \, .
\label{eq:scalarKGexplicit}
\end{equation}
Both particles and antiparticles, with any spin orientation, equally contribute to $f$.

\begin{figure}[hbtp]
\begin{subfigure}{0.33\textwidth}
        \centering
    \includegraphics[width=\textwidth]{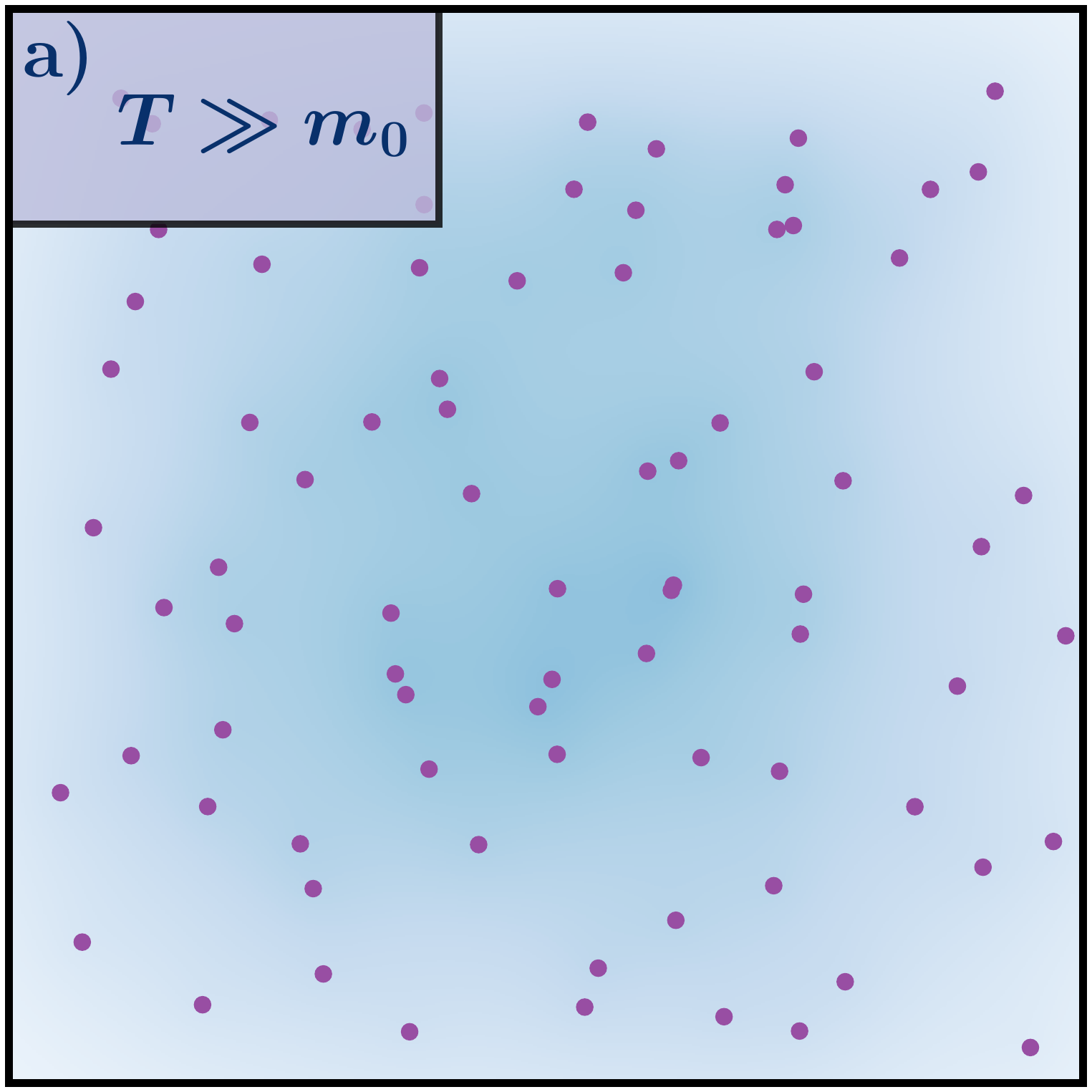}
\end{subfigure}\begin{subfigure}{0.33\textwidth}
        \centering
    \includegraphics[width=\textwidth]{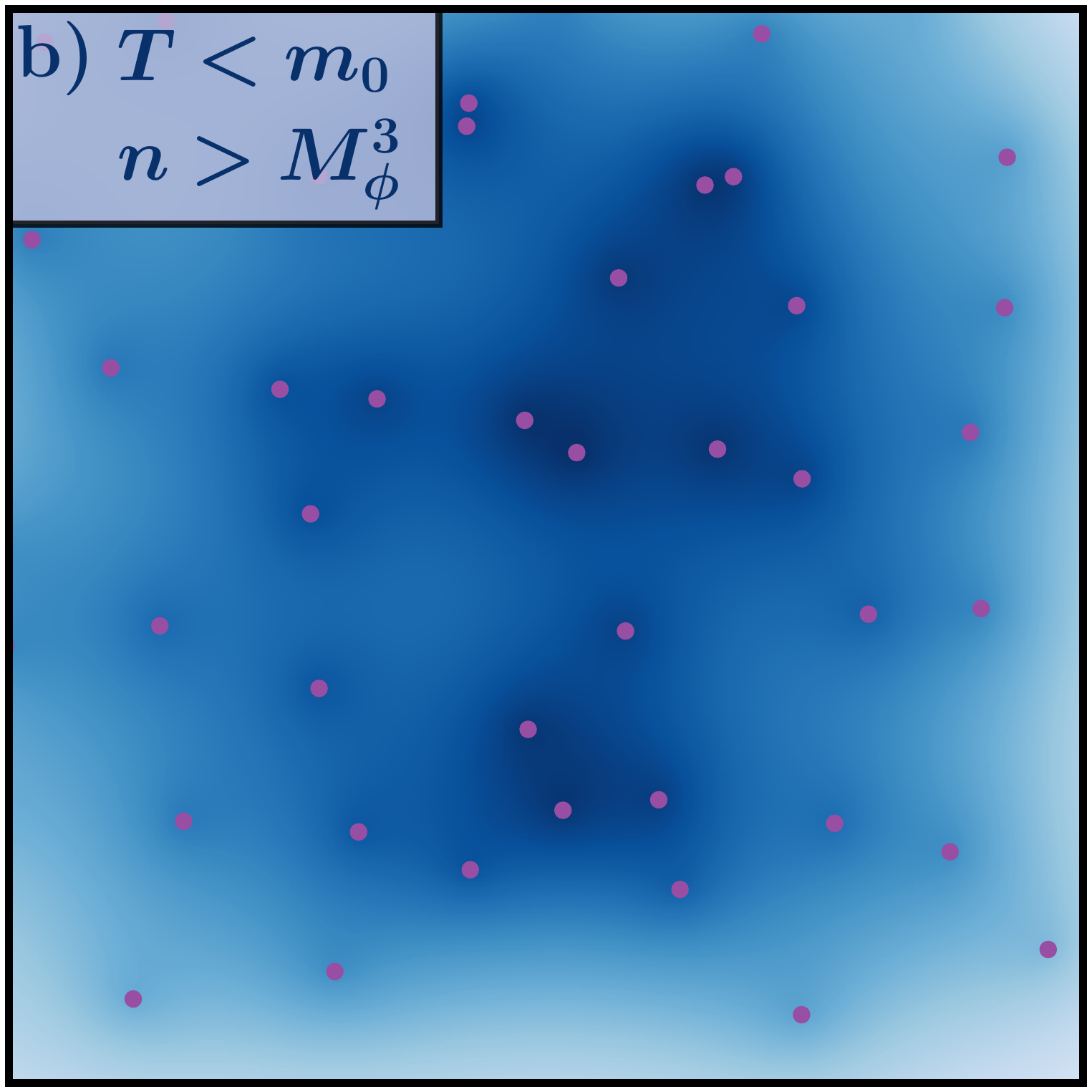}
\end{subfigure}\begin{subfigure}{0.33\textwidth}
        \centering
    \includegraphics[width=\textwidth]{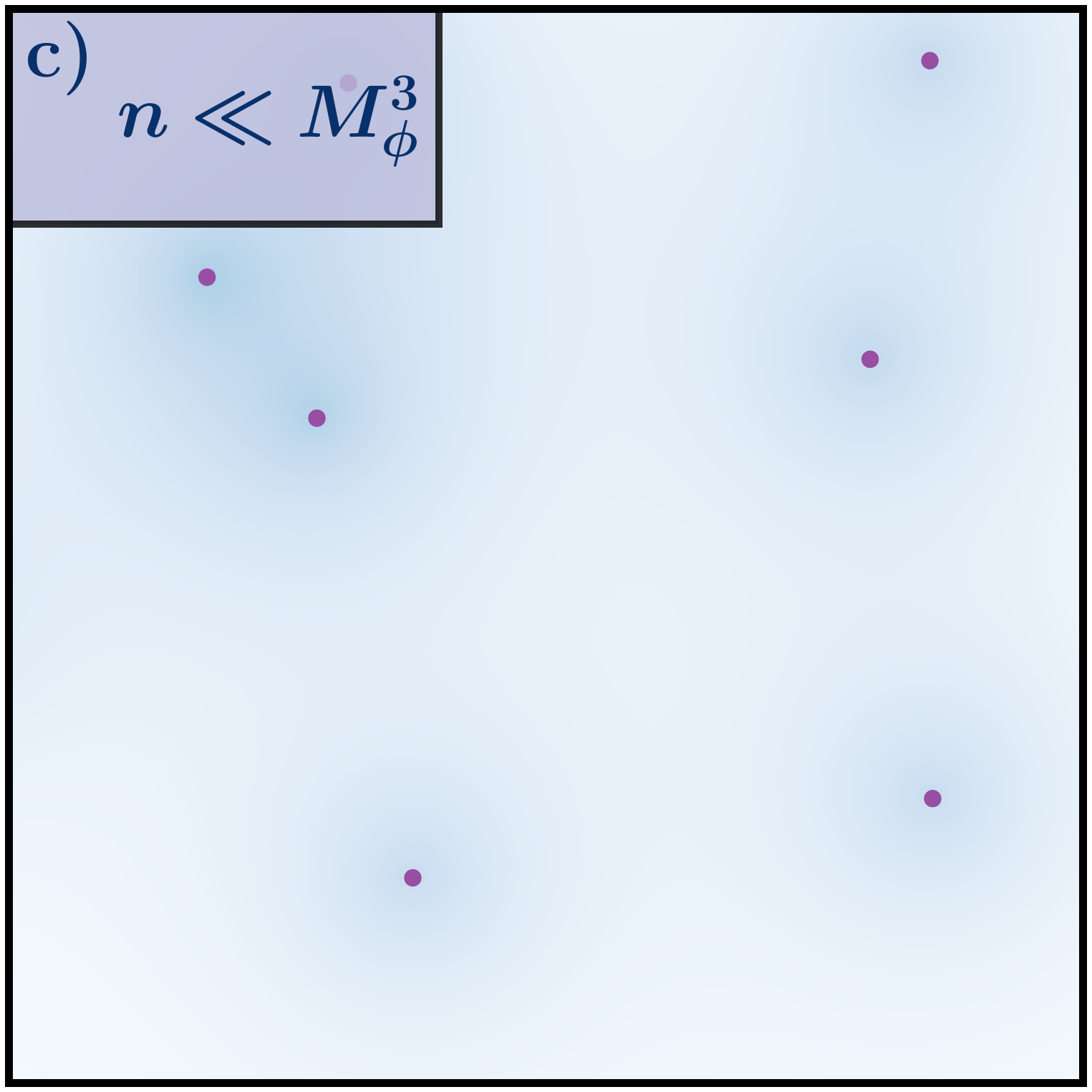}
\end{subfigure}
\caption{Illustration of the three relevant physical regimes for a system of massive 
fermions [purple dots] with a long range interaction mediated by a
scalar field [blue]. $T$ and $n$ are the characteristic fermion
kinetic energy and number density, respectively. For ultrarelativistic fermions, 
in panel (a); and interparticle distances above the interaction range
$\sim M_\phi^{-1}$, in panel (c); there are no long range
effects.}
\label{fig:cartoon}
\end{figure}

\Cref{eq:scalarKGexplicit} shows that the sourced scalar field $\phi$ is generically
suppressed in two scenarios:
\begin{enumerate}
    \item $|\vec{p}| \gg m_0$,\footnote{As we will see in \cref{sec:bkgequations}, for stationary solutions $-m_0 \leq g \phi \leq 0$. Thus, $|\vec{p}| \gg m_0$ also implies $|\vec{p}| \gg m_0 + g \phi$.} i.e., for ultrarelativistic fermions.
    \item $\int f \, d^3\vec{p} \ll M_\phi^3$, i.e., for number densities much smaller than the inverse interaction volume.
\end{enumerate}
This is illustrated in \cref{fig:cartoon}, where we illustrate in blue the scalar field sourced by fermions 
(purple dots) with a characteristic energy $T$ and number density $n$.

The fermion distribution function $f(x^\mu, P_\mu)$ evolves according to the 
Boltzmann equation~\cite{Ma:1995ey}
\begin{equation}
\frac{\partial f}{\partial x^0} + \frac{\mathrm{d}x^i}{\mathrm{d}x^0}\frac{\partial f}{\partial x^i} + \frac{\mathrm{d}P_i}{\mathrm{d}x^0} \frac{\partial f}{\partial P_i} = \left(\frac{\partial f}{\partial x^0}\right)_C \, ,
\label{eq:Boltzmann}
\end{equation}
where the right hand side is the collision term, $\frac{\mathrm{d}x^i}{\mathrm{d}x^0} = \frac{P^i}{P^0}$~\cite{Ma:1995ey}, and 
$\frac{\mathrm{d}P_i}{\mathrm{d}x^0}$ can be obtained from the 
geodesic equation of a fermion coupled with an external scalar 
field~\cite{Ichiki:2007ng,Anderson:1997un}
\begin{equation}
    P^0 \frac{\mathrm{d}P^\mu}{\mathrm{d}x^0} + \Gamma^\mu_{\alpha \beta} P^\alpha P^\beta = - \tilde{m} g \partial^\mu \phi \, ,
\label{eq:geodesic}
\end{equation}
where $\Gamma^\mu_{\alpha \beta}$ are the Christoffel symbols.

\subsection{Homogeneous and Isotropic Scenario}
\label{sec:homogeneous}
\subsubsection{Evolution Equations}
\label{sec:bkgequations}
\Cref{eq:scalarKGexplicit,eq:Boltzmann,eq:geodesic} fully 
characterize the evolution of the system. In this Section, we will
solve them for a spatially flat homogeneous and isotropic 
Universe, described by the FLRW 
metric~\cite{Friedman:1922kd,Friedmann:1924bb,Robertson:1935zz,Walker:1936aa}
\begin{equation}
    \mathrm{d}s^2 = a(\tau)^2 \left(-\mathrm{d}\tau^2 + \delta_{ij} \mathrm{d}x^i \mathrm{d}x^j\right) \, ,
\end{equation}
where $a$ is the scale factor, and $\tau \equiv \int \frac{\mathrm{d}t}{a}$ is the conformal time with $t$ the 
cosmological time. The evolution of the scale factor is related to the total energy density of the Universe $\rho_\mathrm{tot}$ through the Friedman equation
\begin{equation}
    \left(\frac{\mathrm{d}a}{\mathrm{d}\tau}\right)^2 = \frac{8 \pi G_N}{3} a(\tau)^4 \rho_\mathrm{tot} \, ,
\end{equation}
with $G_N$ the Newton gravitational constant.

In a homogeneous and isotropic Universe, the fermion distribution 
function $f(x^\mu, P_\mu)$ can only depend on $\tau$ and the 
modulus of the momentum:
\begin{equation}
  f(x^\mu, P_\mu) = f_0(\tau, q) \, ,
  \label{eq:Boltzmann_bkg_general} 
\end{equation}
where $q \equiv a |\vec{p}|$. The Boltzmann equation then reads
\begin{equation}
  \frac{\partial f_0}{\partial \tau} + \frac{\mathrm{d}q}{\mathrm{d}\tau} \frac{\partial f_0}{\partial q} = \left(\frac{\partial f_0}{\partial \tau}\right)_C \, .
\end{equation}
$\frac{\mathrm{d}q}{\mathrm{d}\tau}$ can be obtained from the 0th
component of the geodesic equation~\eqref{eq:geodesic}. After some
algebra, we obtain
\begin{equation}
  \frac{\mathrm{d}q}{\mathrm{d}\tau} = 0 \, .
\end{equation}

As discussed in the Introduction, we are interested in studying long 
range interaction effects. To isolate them, we will not 
include the collision term in the Boltzmann 
equation~\eqref{eq:Boltzmann_bkg_general}.\footnote{Notice that, if
collisions are relevant, one may also have to consider $\phi$ particle production and fermion-antifermion annihilation.} Physically, this would correspond 
to small coupling constants $g$. As we will see, long range 
effects can still be relevant since they scale as 
$\frac{g}{M_\phi}$. Under this hypothesis, \cref{eq:Boltzmann_bkg_general} reads
\begin{equation}
    \frac{\partial f_0(\tau, q)}{\partial \tau} = 0 \, .
    \label{eq:Boltzmann_bkg} 
\end{equation}
That is, \emph{any homogeneous and isotropic fermion distribution function that depends only 
on the combination $q = a |\vec{p}|$ does not evolve with time even 
under the presence of a long range scalar interaction}.

The equation of motion~\eqref{eq:scalarKGexplicit} for a
homogeneous scalar field $\phi_0(\tau)$ reads
\begin{equation}
\frac{\phi_0''}{a^2} + 2 H \frac{\phi_0'}{a} + M_\phi^2 \phi_0 = - g \int \mathrm{d}^3 \vec{p} \frac{m_0 + g \phi_0}{\sqrt{|\vec{p}|^2 + (m_0 + g \phi_0)^2}} f_0(\tau, a|\vec{p}|) \, ,
\label{eq:scalarKGbackground}
\end{equation}
where primes denote derivatives with respect to conformal time and  
$H \equiv \frac{1}{a}\frac{\mathrm{d}a}{\mathrm{d}t} = \frac{a'}{a^2}$
is the Hubble parameter. That is, we obtain a Klein-Gordon equation
with a field-dependent source term, which will induce an effective scalar mass
\begin{equation}
    M_T^2 \equiv \frac{\partial}{\partial \phi_0} \left(g \int \mathrm{d}^3 \vec{p} \frac{m_0 + g \phi_0}{\sqrt{|\vec{p}|^2 + (m_0 + g \phi_0)^2}} f_0(\tau, a|\vec{p}|) \right) \, .
    \label{eq:thermalMass}
\end{equation}
%If M<<H and MT<<H, quintessence
%If M>>H and MT>>H, our setup
%Try to introduce MT from the beginning

In Eq.~\eqref{eq:scalarKGbackground}, there are two characteristic 
timescales: on the one hand, $H^{-1}$, which controls both the Hubble friction term $2 H \frac{\phi_0'}{a}$ as well as the 
rate at which the right-hand side changes.\footnote{Because of \cref{eq:Boltzmann_bkg}, $f_0(\tau, a|\vec{p}|)$ only depends on time through the scale factor in the second argument.} On the other hand, the inverse scalar field mass 
\begin{equation}
    M_\mathrm{eff}^{-1} \equiv (M_\phi^2 + M_T^2)^{-1/2} \, ,
    \label{eq:scalarMeff}
\end{equation}
which controls its characteristic oscillation time. Depending on the relative values 
of these timescales, we can distinguish three qualitatively different scenarios:
\begin{itemize}
    \item $M_\mathrm{eff} \ll H$ for all relevant times. In this case, $\phi_0(\tau)$ would be frozen to its value after inflation, and the physics of the scalar field would be that of quintessence, widely studied in the literature~\cite{Ratra:1987rm, Wetterich:1987fm, Frieman:1995pm, Coble:1996te, Caldwell:1997ii, Copeland:2006wr, Tsujikawa:2013fta, DAmico:2018hgc}. Furthermore, the scalar field sourced by the fermion background (that is, the right-hand side in Eq.~\eqref{eq:scalarKGbackground}) would play no significant role. Since we are interested in the effect of fermion self interactions, we will not consider this scenario in this work.
    
    \item $M_\mathrm{eff} \sim H$. In this case, $\phi_0(\tau)$ is determined by a non-trivial interplay among its initial condition and the fermion background. As we only want to study the effect of the latter, we will not consider this scenario in this work.
    
    \item $M_\mathrm{eff} \gg H$ for all relevant times. As we will see next, in this case the physics of a fermion background interacting with a scalar field is insensitive to the initial condition of the latter. This will be the scenario studied in this work.
\end{itemize} %Later: say we want M>>H because we want to study the physics insensitive to initial condition
We can study the $M_\mathrm{eff} \gg H$ scenario by using the adiabatic approximation. This corresponds 
to writing $\phi_0(\tau) = \overline{\phi_0}(\tau) + \varphi(\tau)$, where $\overline{\phi_0}$ satisfies
\begin{equation}
M_\phi^2 \overline{\phi_0} \equiv - g \int \mathrm{d}^3 \vec{p} \frac{m_0 + g \overline{\phi_0}}{\sqrt{|\vec{p}|^2 + (m_0 + g \overline{\phi_0})^2}} f_0(\tau, a |\vec{p}|) \, .
\label{eq:scalarBackground}
\end{equation}
The evolution equation for $\varphi$ then reads
\begin{equation}
\frac{\varphi''}{a^2} + 2 H \frac{\varphi'}{a} + \left(M^2 + \overline{M_T^2}\right) \varphi + \mathcal{O}(\varphi^2) = \mathcal{O}\left(H^2 \overline{\phi_0}\right) \, ,
\label{eq:scalarOscillations}
\end{equation}
where $\overline{M_T^2}$ is given by \cref{eq:thermalMass} evaluated for $\phi_0 = \overline{\phi_0}$. That is, the scalar field separates into a component sourced by the 
fermions, \cref{eq:scalarBackground}; and a fastly oscillating component, 
satisfying \cref{eq:scalarOscillations}. The latter corresponds to a background of $\phi$ particles 
at rest, and it is nonzero only if set by the initial condition (up to small corrections $\mathcal{O}
\left(\frac{H^2}{M^2 + M_T^2} \right)$). Since, on top of that, it does not affect the scalar field 
sourced by the fermions $\overline{\phi_0}$, we will not study it. In
what follows, to simplify notation $\phi_0$ will refer to
$\overline{\phi_0}$, and $M_T^2$ to $\overline{M_T^2}$.

\subsubsection{Solution for a Thermal Fermion Relic}

To compute the scalar field $\phi_0$ and obtain the macroscopic properties of the system, 
we have to specify the fermion distribution function $f_0$. To this purpose, we will assume that 
the fermions were in the past in thermal equilibrium. As we are neglecting 
collisions, they must thermally decouple before long range effects become relevant, i.e., while 
still relativistic. In this case, we can assume that the fermion distribution function takes a Fermi-Dirac form,
\begin{equation}
f_0(\tau, a|\vec{p}|) = \frac{\mathfrak{g}}{(2 \pi)^3} \frac{1}{e^{|\vec{p}|/T(a)} + 1} \, .
\label{eq:bkgFD}
\end{equation}
Here, $\mathfrak{g}$ is the amount of internal degrees of freedom of the fermion (including 
particles, antiparticles and any internal quantum number) and $T$ its temperature. The Boltzmann 
equation~\eqref{eq:Boltzmann_bkg} then requires $T \propto \frac{1}{a}$. This distribution applies, 
e.g., to neutrinos and other hot thermal relics. 
Exceptions include particles that never reach thermal equilibrium (e.g., produced through 
freeze-in) or non-negligible chemical potentials.

Using \cref{eq:bkgFD}, we can self-consistently solve \cref{eq:scalarBackground} to obtain the 
scalar field $\phi_0$ sourced by the fermions. Then, we can compute the effective fermion mass 
$\tilde{m} = m_0 + g \phi_0$ as well as the energy density $\rho$ and
pressure $P$ of the system
\begin{align}
    \rho &= \rho_\phi + \rho_F = \frac{1}{2} M_\phi^2 \phi_0^2 +  \int \mathrm{d}^3\vec{p} \, \sqrt{|\vec{p}|^2 + \tilde{m}^2} f_0(\tau, a|\vec{p}|) \, , \label{eq:rhobkg}\\
    P &= P_\phi + P_F = -\frac{1}{2} M_\phi^2 \phi_0^2 +  \int
    \mathrm{d}^3\vec{p} \, \frac{|\vec{p}|}{3 \sqrt{|\vec{p}|^2 +
        \tilde{m}^2}} f_0(\tau, a|\vec{p}|) \, . \label{eq:pbkg}
\end{align}
We have checked that, under our assumption $M_\phi^2 + M_T^2 \gg H^2$, the kinetic term 
$\frac{1}{2}\dot{\phi}_0^2$ in $\rho_\phi$ and $P_\phi$ can be neglected.

\begin{figure}[hbtp]
\centering
\begin{subfigure}[t]{0.48\textwidth}
        \centering
    \includegraphics[width=\textwidth]{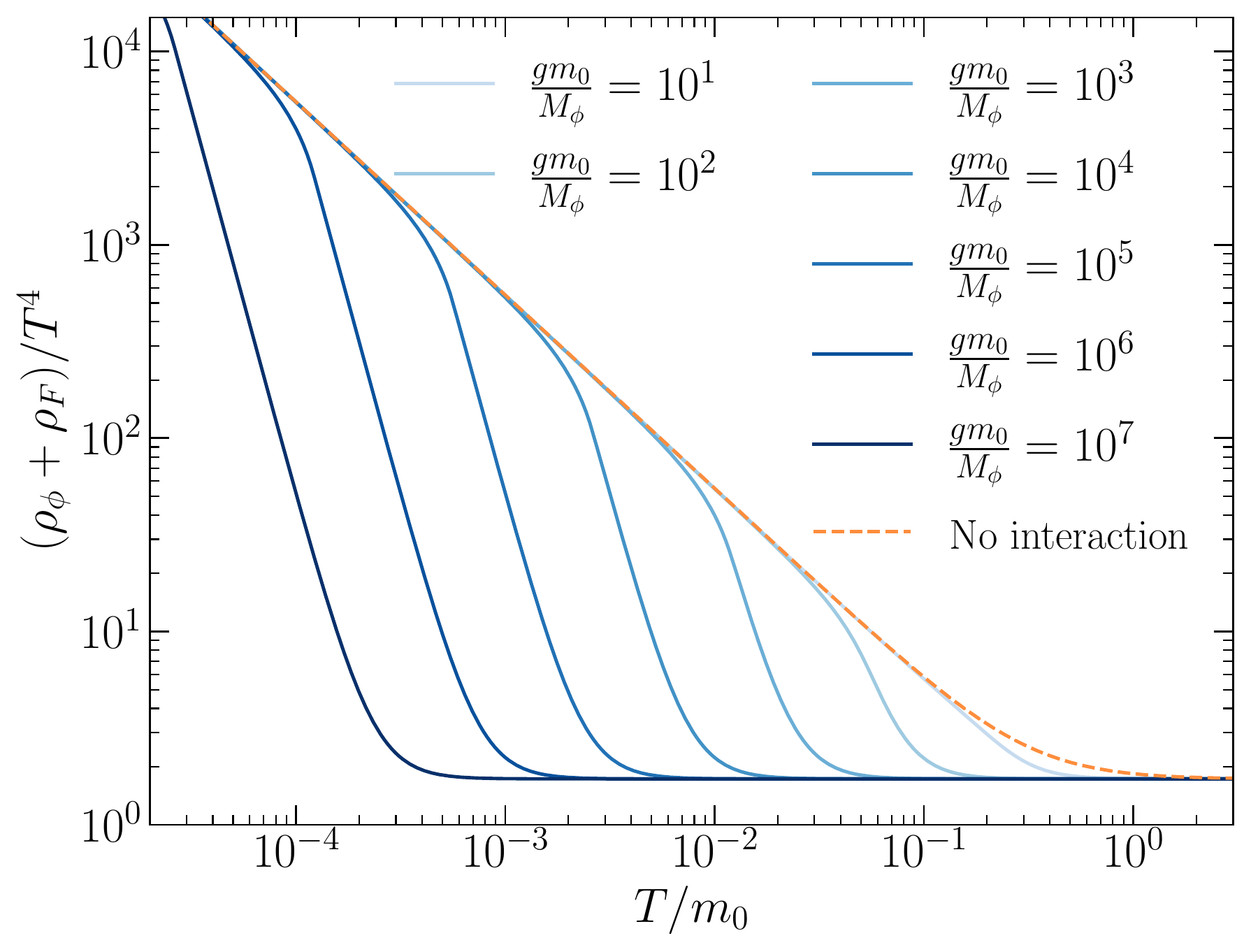}
    \caption{Energy density of the system divided by $T^4$.}
    \label{fig:backgroundA}
\end{subfigure}\hspace{0.04\textwidth}\begin{subfigure}[t]{0.48\textwidth}
        \centering
    \includegraphics[width=\textwidth]{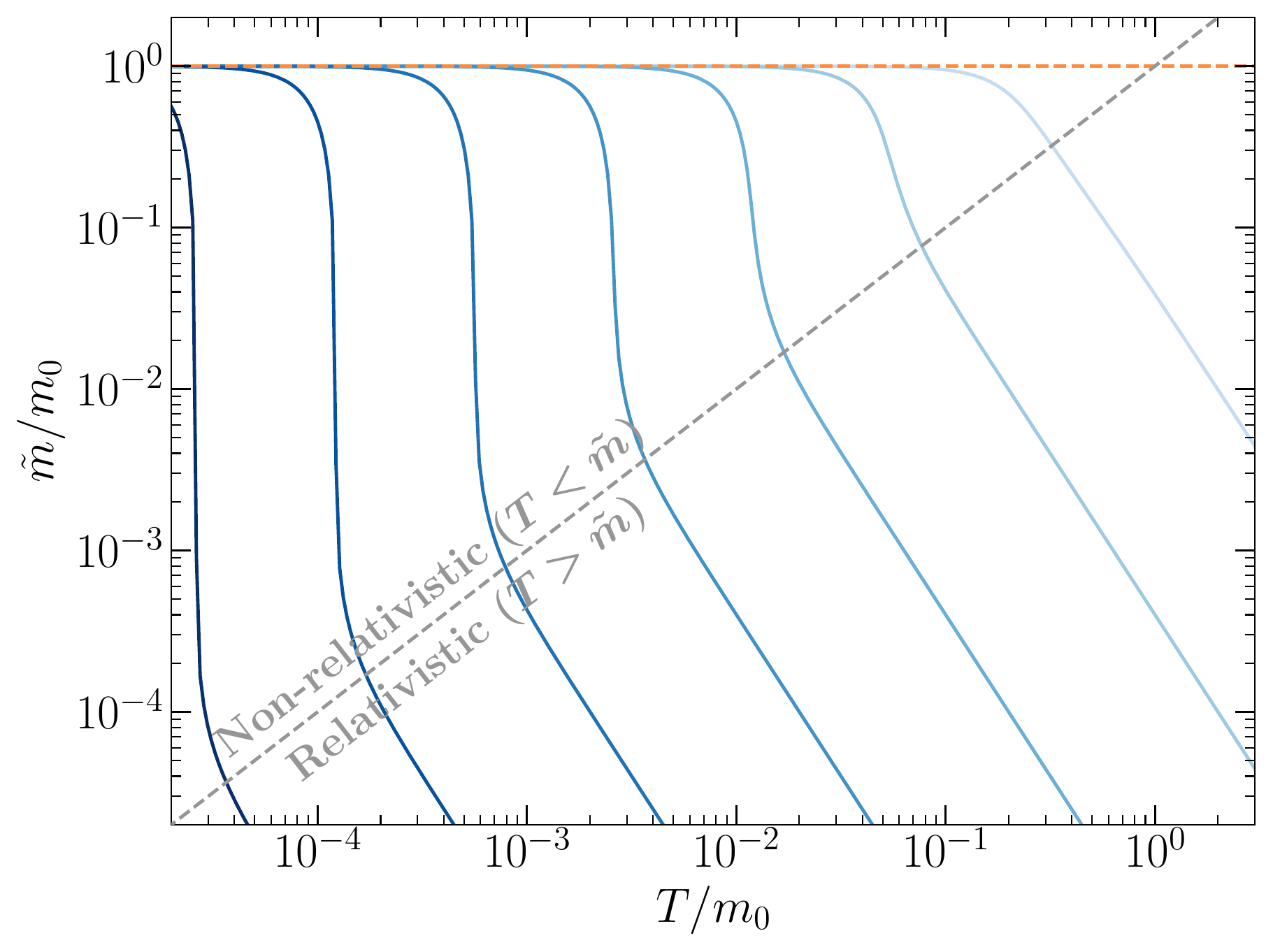}
    \caption{Effective fermion mass divided by $m_0$. The gray line separates the non-relativistic and relativistic regimes.}
    \label{fig:backgroundB}
\end{subfigure}
\caption{Energy density and effective fermion mass as a function of $T/m_0$ for different interaction strengths [solid blue shades]. The dashed orange line shows the result without long range interactions. $T$ is the fermion temperature, $m_0$ its vacuum mass, and $\frac{g m_0}{M_\phi}$ parametrizes the interaction strength. The fermion distribution function is given by \cref{eq:bkgFD} with $\mathfrak{g}=6$ degrees of freedom.}
\label{fig:background}
\end{figure}

From \cref{eq:scalarBackground}, one can see that $g$ and $M_\phi$ only enter into the 
homogeneous and isotropic results through the combination $\frac{g m_0}{M_\phi}$. Thus, 
we show in \cref{fig:background} for different values of $\frac{g m_0}{M_\phi}$ the 
energy density $\rho$ (normalized to $T^4$) as well as the effective fermion mass 
$\tilde{m}$ (normalized to its vacuum mass $m_0$) as a function of $T$ (normalized to  
$m_0$). We have chosen $\mathfrak{g} = 6$, so that our results directly apply
to three interacting neutrino and antineutrino species.

\Cref{fig:background} allows to understand the cosmological evolution of the system. In the very early Universe 
($T \gg m_0$) all fermions are ultrarelativistic and, as seen in the 
rightmost side of \cref{fig:backgroundA}, the energy density of the system is that of a non-interacting 
gas of ultrarelativistic fermions. As the Universe expands, the temperature decreases and for strong enough interactions the sourced scalar field reduces the effective fermion mass (see 
\cref{fig:backgroundB}), keeping fermions relativistic even though $T \ll m_0$. Therefore, they will 
contribute to the energy density as $\rho_F \propto T^4$. As we will see, the scalar field contribution 
is independent of $T$, so at some point it takes over the fermion contribution, giving the steeply 
increasing energy density seen in \cref{fig:backgroundA}. Finally, when 
the temperature is small enough, the interparticle distance is larger than the interaction range and all 
interaction effects switch off: the energy density is that of a non-interacting gas of fermions, and 
$\tilde{m} = m_0$.

\begin{figure}[hbtp]
    \centering
    \includegraphics[width=0.85\textwidth]{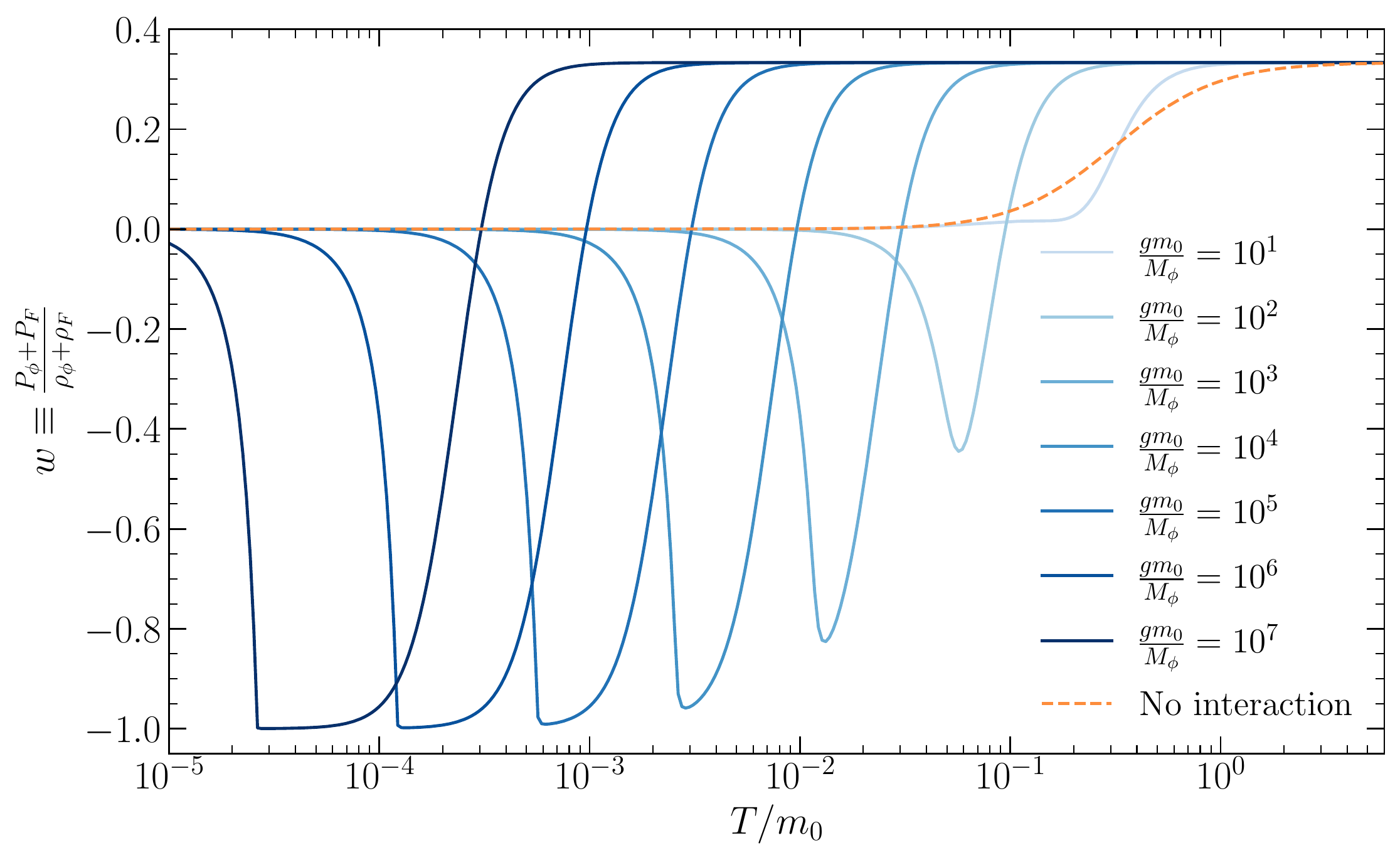}
    \caption{Equation of state of the system as a function of $T/m_0$ for different interaction strengths [solid blue shades]. The dashed orange line shows the result without long range interactions. $T$ is the temperature, $m_0$ the vacuum fermion mass, and $\frac{g m_0}{M_\phi}$ parametrizes the interaction strength. The fermion distribution function is given by \cref{eq:bkgFD} with $\mathfrak{g}=6$ degrees of freedom.}
    \label{fig:eos}
\end{figure}

The rate at which the energy density changes can be quantified through the equation of state parameter 
$w \equiv P/\rho$, as in an expanding Universe $\frac{1}{\rho} \frac{\mathrm{d}\rho}{\mathrm{d}t} = - 3 H (1 + w)$. This is shown in \cref{fig:eos} as a function of the fermion temperature $T$ (normalized 
to its vacuum mass $m_0$) for different interaction strengths. As we can see, for $T \gg m_0$ there are no 
interaction effects and the equation of state is that of an ideal gas of non-interacting relativistic 
fermions, $w = 1/3$.  As the temperature decreases, the interaction keeps fermions 
ultrarelativistic, and $w$ is still $1/3$ until the scalar field energy density and pressure take over 
those of the fermions. At this point, the equation of state parameter can take negative values, even 
reaching $w \simeq -1$. Finally, as the system cools down the interparticle distance gets larger than 
the interaction range and all interaction effects switch off. The equation of state is then that of a 
non-interacting gas of non-relativistic fermions, $w=0$.

The behavior in \cref{fig:background,fig:eos} can be analytically understood by approximately solving 
\cref{eq:scalarBackground}. If $T \gg \tilde{m}$, we can neglect the second term in the square root and
\begin{align}
    \phi_0 & \simeq - \frac{ \frac{\mathfrak{g}}{24} g \frac{m_0}{T} T^3}{M_\phi^2 + \frac{\mathfrak{g}}{24} g^2 T^2} = - \frac{g}{M_\phi^2} \frac{\tilde{m}}{T} \frac{\mathfrak{g}}{24} T^3 \, , \\
    \tilde{m} & \simeq m_0 \frac{1}{1 + \frac{\mathfrak{g}}{24} \frac{g^2 T^2}{M_\phi^2}} \, . \label{eq:mtildeapprox}
\end{align}
In this limit, the scalar field $\phi_0$ is simply proportional to the product of the coupling $g$; a 
factor $\frac{m_0}{T}$ that, as we anticipated, suppresses long range effects for $T \gg m_0$; and 
$T^3$, proportional to the fermion number density. In the denominator, we have the effective scalar mass 
squared $M_\mathrm{eff}^2 \equiv M_\phi^2 + M_T^2 \simeq M_\phi^2 + \frac{\mathfrak{g}}{24} g^2 T^2$, 
 that enhances $\phi_0$ for longer interaction ranges 
(i.e., smaller scalar masses). Notice that the naive enhancement of the interaction by reducing $M_\phi$ 
to increase its range is only effective until $M_\phi \lesssim
M_T$. Finally, from \cref{eq:mtildeapprox} we see
that as long as $\frac{\mathfrak{g}}{24} \frac{g^2 T^2}{M_\phi^2} \gg \frac{m_0}{T}$, the sourced scalar 
field keeps the fermions ultrarelativistic.

We can also analytically understand why in \cref{fig:eos} there is a region with a dark energy-like 
equation of state, $w \simeq -1$. Using \cref{eq:mtildeapprox}, we can write $\phi_0 \propto 
\frac{\tilde{m}}{T} T^3$. For high enough temperatures, $\frac{\tilde{m}}{T} \sim \frac{1}{T^3}$, and so 
the scalar field stays constant as the Universe expands. In other words, the decrease of long range 
effects due to the dilution of the fermions is exactly compensated by them becoming less relativistic.

On the other hand, for $T \ll \tilde{m}$ (which, as we have seen above, requires $\frac{\mathfrak{g}}{24} \frac{g^2 T^2}{M_\phi^2} \lesssim \frac{m_0}{T}$) ,
\begin{align}
\phi_0 & \simeq - \frac{3 \zeta(3) \mathfrak{g}}{4\pi^2} g \frac{T^3}{M_\phi^2} \, , \\
\tilde{m} & \simeq m_0 \left(1 - \frac{3 \zeta(3) \mathfrak{g}}{4\pi^2} \frac{g^2 T^2}{M^2} \frac{T}{m_0}\right) \, .
\end{align}
That is, $\phi_0$ is the product of the coupling and the fermion number 
density divided by the vacuum scalar mass squared (as in this limit $M_T$ is negligible). 
As the temperature decreases, the scalar field energy density dilutes
as $T^6$, whereas the fermion energy
density dilutes slower, as $T^3$.  At the same time, fermions rapidly
acquire their vacuum mass. In other words, all the long range effects rapidly turn off as intuitively expected from an interaction whose energy 
density is proportional to the fermion number density squared.

\subsection{Perturbations and Instability}
\label{sec:perturbations}
After having discussed the evolution of a homogeneous and isotropic background, the next step to 
characterize the cosmology of a system of long range interacting fermions is to study linear 
inhomogeneous perturbations. In the following, we will work in the synchronous gauge~\cite{Ma:1995ey}
\begin{equation}
    \mathrm{d}s^2 = a(\tau)^2 [- \mathrm{d}\tau^2 + (\delta_{ij} + h_{ij}(\vec{x},\tau)) \mathrm{d}x^i \mathrm{d}x^j] \, .
\end{equation}
We will only consider scalar metric perturbations, that can be Fourier expanded as
\begin{equation}
    h_{ij}(\vec{x},\tau) = \int \mathrm{d}^3k \, e^{i \vec{k}\cdot\vec{x}} \left[\hat{k}_i \hat{k}_j h(\vec{k}, \tau) + 
    \left(\hat{k}_i \hat{k}_j - \frac{1}{3}\delta_{ij}\right) 6 \eta(\vec{k},\tau)\right] \, ,
\end{equation}
where $\hat{k} \equiv \frac{\vec{k}}{|\vec{k}|}$ and $h(\vec{k},\tau)$ and $\eta(\vec{k},\tau)$ are the 
scalar metric perturbations in Fourier space. In addition, we will write the fermion distribution 
function and scalar field as
\begin{align}
f(x^\mu, P_\mu) & = f_0(\tau, q) \left[ 1 + \Psi(\vec{x}, \tau, q, \hat{n})\right] \, , \\
\phi(x^\mu) & = \phi_0(\tau) + \delta \phi(\vec{x}, \tau) \, ,
\end{align}
where $\hat{n} \equiv \frac{\vec{p}}{|\vec{p}|}$.

The Boltzmann equation~\eqref{eq:Boltzmann} in Fourier space reads, to
linear order in perturbations,
\begin{equation}
\Psi' + i \frac{q}{\varepsilon} (\vec{k} \cdot \hat{n}) \Psi + \frac{\mathrm{d} \log f_0}{\mathrm{d}\log q} \left[\eta' - \frac{h' + 6 \eta'}{2} (\hat{k} \cdot \hat{n})^2 - g \frac{a^2 \tilde{m}}{q \varepsilon} i (\vec{k} \cdot \hat{n}) \delta \phi\right] = 0 \, ,
\label{eq:BoltzmannPerturb}
\end{equation}
where $\epsilon \equiv \sqrt{q^2 + \tilde{m}^2 a^2}$. The term in square brackets corresponds to the 
effect of long range interactions, either gravitational, parametrized by $\eta$ and $h$; or induced by 
the scalar field, proportional to $g \, \vec{p} \cdot \vec{\nabla} \delta \phi \sim g (\hat{k} \cdot \hat{n}) \delta \phi$ and suppressed by $\frac{\tilde{m}}{\varepsilon}$ for ultrarelativistic fermions. 
Finally, we can expand $\Psi$ in Legendre polynomials following the
conventions in Ref.~\cite{Ma:1995ey}, obtaining the following tower of
Boltzmann equations
\begin{align}
\Psi_0' & = -\frac{q k}{\varepsilon} \Psi_1 + \frac{1}{6} h' \frac{\mathrm{d} \log f_0}{\mathrm{d}\log q} \, , \\
\Psi_1' & = \frac{q k}{3\varepsilon} (\Psi_0 - 2 \Psi_2) -  g \frac{a^2 \tilde{m} k}{3q \varepsilon} \delta \phi \frac{\mathrm{d} \log f_0}{\mathrm{d}\log q} \, , \\
\Psi_2' & = \frac{q k}{5\varepsilon} (2 \Psi_1 - 3 \Psi_3) - \left(\frac{1}{15} h' + \frac{2}{5} \eta' \right) \frac{\mathrm{d} \log f_0}{\mathrm{d}\log q} \, , \\
\Psi_\ell' & = \frac{q k}{(2\ell+1)\varepsilon} [\ell \Psi_{\ell-1} - (\ell+1) \Psi_{\ell+1}] \quad \forall \ell \geq 3 \, .
\end{align}
That is, the effect of long range interactions enters both through a time-dependent mass $\tilde{m}$ as 
well as through an interaction with scalar field perturbations in the $\ell=1$ multipole.

Regarding the equation for the scalar
field~\eqref{eq:scalarKGexplicit}, it reads in Fourier space to linear order
\begin{equation}
    \delta \phi'' + 2 a H \delta \phi' + \frac{1}{2} h' \phi' + \left[k^2 + a^2\left(M_\phi^2 + M_T^2\right)\right] \delta \phi = - g \, 4\pi \int \mathrm{d}q \, q^2\frac{\tilde{m}}{\varepsilon} f_0(q) \Psi_0(q, k, \tau) \, .
\label{eq:perturbedKG}
\end{equation}
As in \cref{sec:bkgequations}, if the effective inverse scalar mass ${\left[(k/a)^2 + M_\phi^2 + M_T^2\right]^{-1/2}}$ is much smaller 
than other timescales, we can apply the adiabatic approximation and
\begin{equation}
    \left[(k/a)^2 + M_\phi^2 + M_T^2\right] \delta \phi \simeq - g \frac{4\pi}{a^2} \int \mathrm{d}q \, q^2\frac{\tilde{m}}{\varepsilon} f_0(q) \Psi_0(q, k, \tau) \, .
    \label{eq:perturbedAdiabatic}
\end{equation}

From \cref{eq:BoltzmannPerturb,eq:perturbedAdiabatic}, we see that the scalar interaction will introduce 
a new attractive long range force among fermions. If it is stronger than or comparable to gravity, it
can significantly affect structure growth, as first pointed out in
Ref.~\cite{Afshordi:2005ym} in the context of neutrino-induced dark
energy models (see also 
Refs.~\cite{Beca:2005gc,Kaplinghat:2006jk,Bjaelde:2007ki,Bean:2007nx}).

In particular, Ref.~\cite{Afshordi:2005ym} found that for scalar masses $M_\mathrm{eff} \gg H$, as in our case,  
non-relativistic fermion density perturbations of sizes $\gtrsim M_\mathrm{eff}^{-1}$ exponentially grow 
over timescales $\ll M_\mathrm{eff}^{-1}$ much shorter than cosmological times. As a consequence, when 
becoming non-relativistic, all fermions will collapse into non-linear structures or \emph{nuggets} with 
typical sizes $< M_\mathrm{eff}^{-1}$, separated by distances $\gg M_\mathrm{eff}^{-1}$. The outcome of 
this transition will be a dilute gas of non-interacting
\emph{nuggets}, with sizes much smaller than cosmological scales,
behaving as dust.

\begin{figure}[hbtp]
    \centering
    \includegraphics[width=0.7\textwidth]{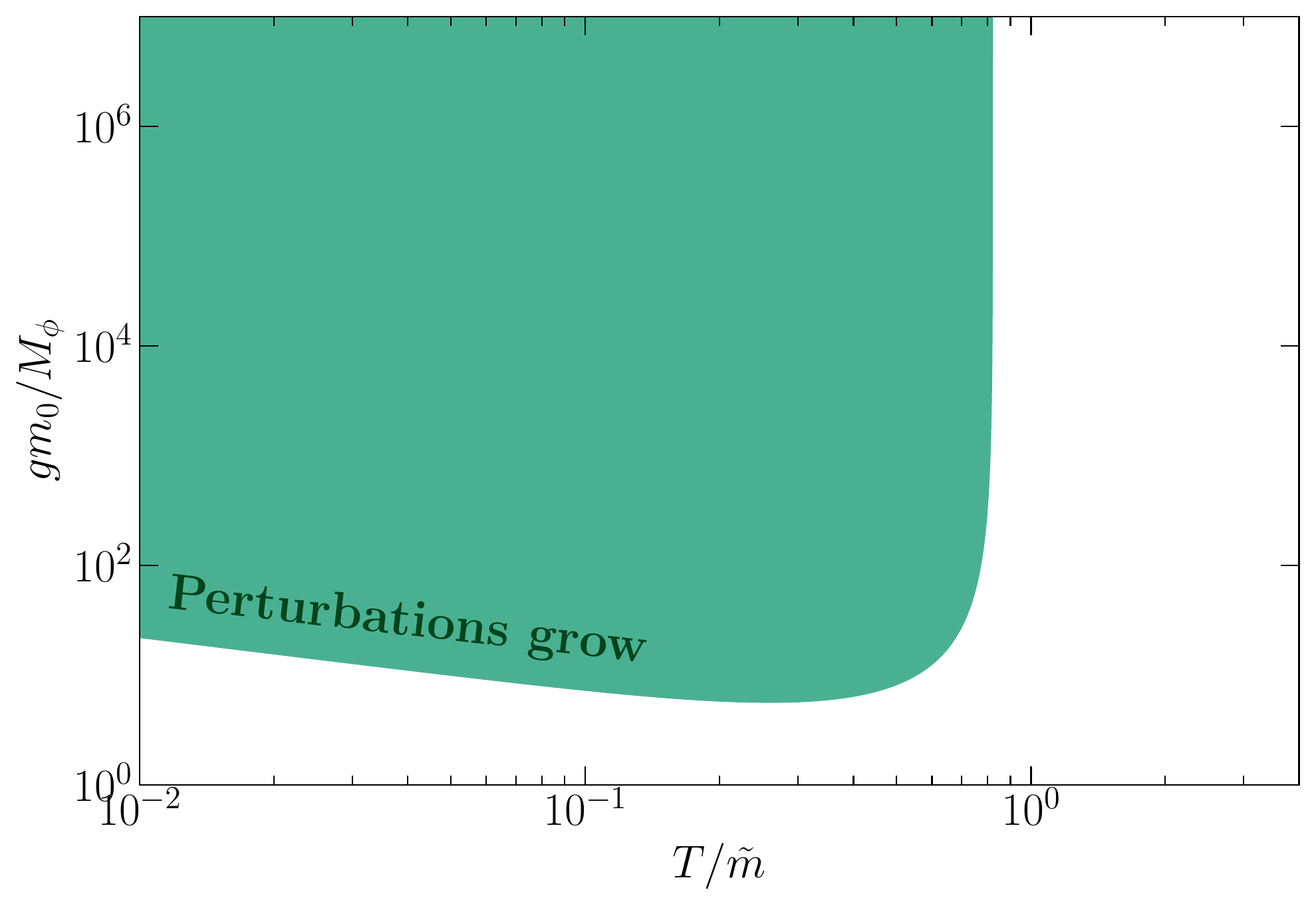}
    \caption{The green region shows the values of the interaction strength $\frac{g m_0}{M_\phi}$ and temperature $T$ (in units of the effective fermion mass $\tilde{m}$) where perturbations grow due to the long range interaction. For large temperatures, relativistic motion inhibits perturbation growth. Below the green region, perturbations do not grow because the relevant interaction range is shorter than the average interparticle distance. The fermion distribution function is given by \cref{eq:bkgFD} with $\mathfrak{g}=6$ degrees of freedom.}
    \label{fig:instability}
\end{figure}

Following Ref.~\cite{Bjaelde:2007ki}, we have computed the interaction strengths in our model for which 
fermion density perturbations exponentially grow (see \cref{sec:app_instability} for the details). Our 
results are depicted in \cref{fig:instability}, where we show in
shaded the values of $\frac{g m_0}{M_\phi}$ and temperature 
(in units of the effective fermion mass $\tilde{m}$) where this
instability is present. As we see, for interactions strengths $\frac{g m_0}{M_\phi} \gtrsim 5$, as soon as fermions
become non-relativistic ($T \lesssim 0.8 \tilde{m}$) the long range interaction makes perturbations 
quickly grow. For $T > \tilde{m}$, perturbations do not grow due to two effects: on the one hand, the 
large dispersion velocities of relativistic particles inhibit perturbation growth; on the other hand, 
scalar interactions are suppressed for relativistic fermions. Notice that the second effect is 
characteristic of our model and is not generically present for other interactions. In addition, for small 
$\frac{g m_0}{M_\phi}$, when fermions become non-relativistic the interparticle distance is larger than 
the interaction range and the scalar self interaction does not induce perturbation growth.

\begin{figure}[hbtp]
\begin{subfigure}{0.5\textwidth}
        \centering
    \includegraphics[width=\textwidth]{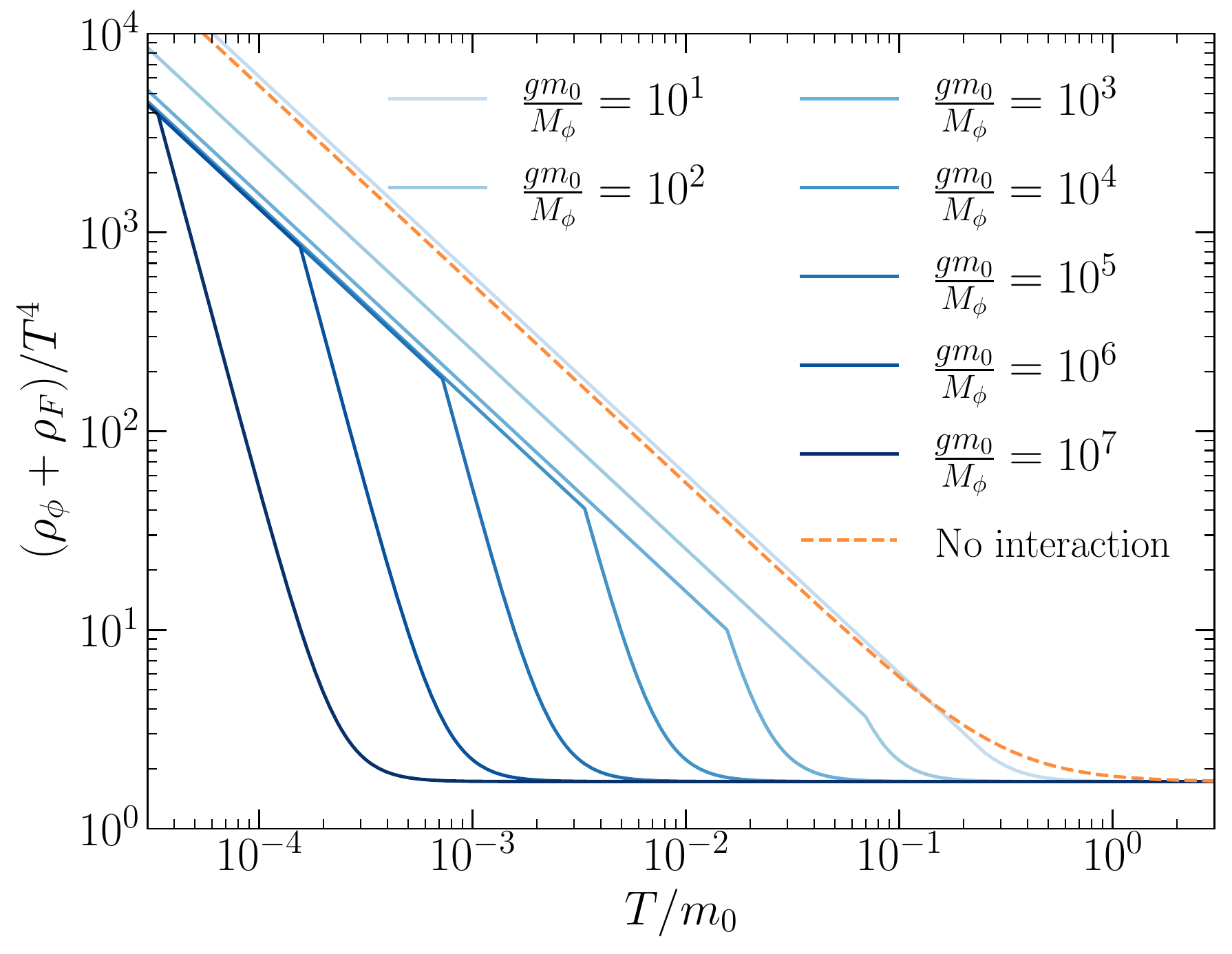}
\end{subfigure}\begin{subfigure}{0.5\textwidth}
        \centering
    \includegraphics[width=\textwidth]{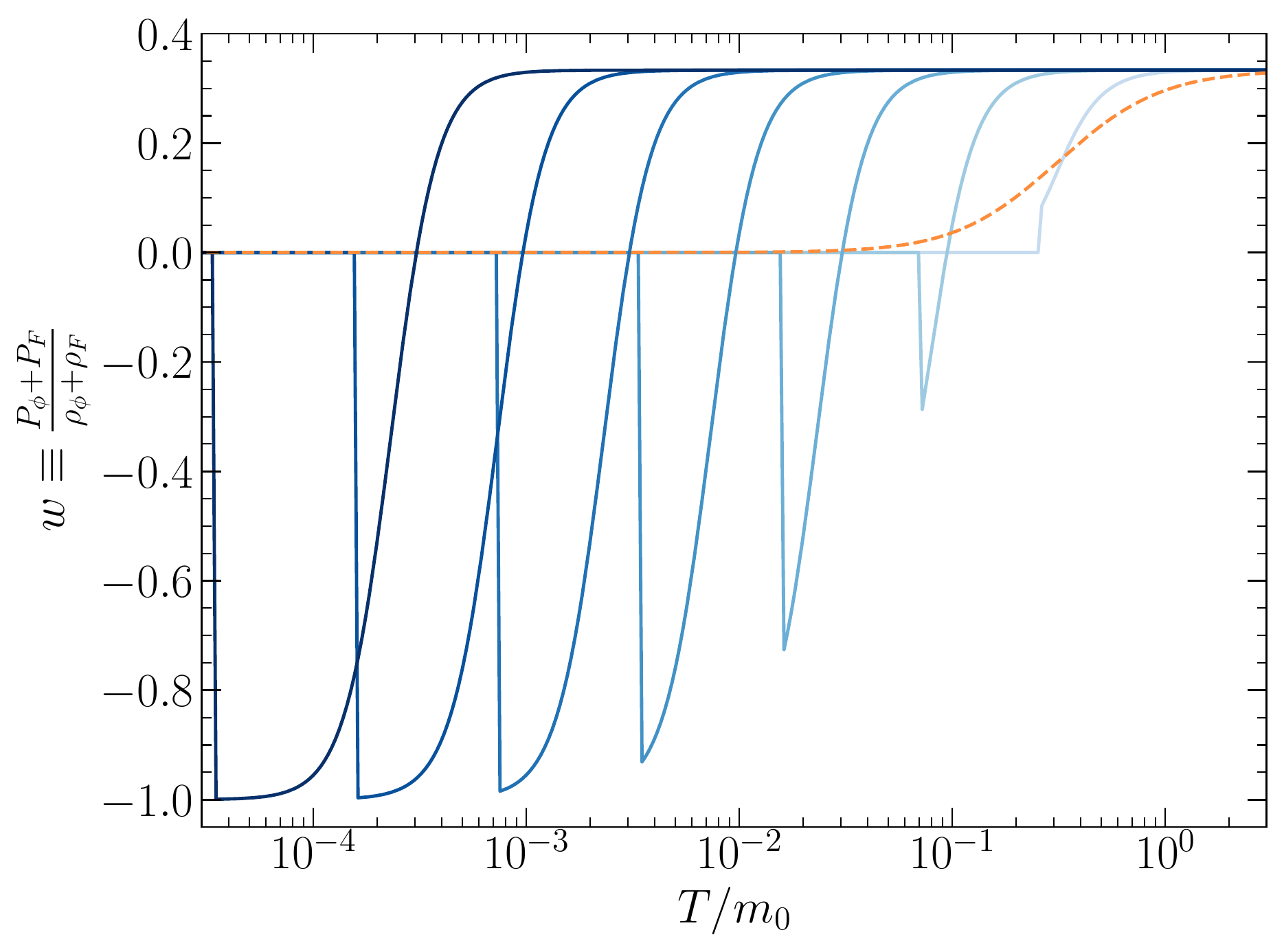}
\end{subfigure}
\caption{Energy density [left] and equation of state [right] of the system for different interactions strengths [solid blue shades], including instantaneous \emph{nugget} formation. See \cref{fig:background,fig:eos} and text for details. The fermion distribution function is given by \cref{eq:bkgFD} with $\mathfrak{g}=6$ degrees of freedom.}
\label{fig:ew_nuggets}
\end{figure}

In order to model this instability, we quantify in \cref{sec:app_instability} the timescale over which 
fermion density perturbations become non-linear due to the exponential growth. This timescale is much 
smaller than cosmological scales as long as
\begin{equation}
    \sqrt{M_\phi^2 + M_T^2} \gtrsim 10^5 H \, .
    \label{eq:nugget_condition}
\end{equation}
We will impose this condition, and consider that as soon as the temperature drops below the unstable 
temperature in \cref{fig:instability},\footnote{We have checked that the final results are not sensitive 
to the specific temperature at which the system becomes unstable within $\lesssim 10\%$ variations of the 
latter.} the system undergoes an instantaneous transition to a dust-like
behavior. Under this assumption, the energy density and equation of state of the fermion background as a 
function of temperature\footnote{Technically, the fermion temperature is not well defined after 
\emph{nugget} formations, as the fermion background does not have a thermal distribution. In this region, 
$T$ should be understood as a proxy for the scale factor.} are shown in \cref{fig:ew_nuggets}. As we see, 
the transition takes place relatively late and most of the phenomenology described in 
\cref{sec:homogeneous} is still valid. Furthermore, the low temperature energy density now depends on the 
interaction strength, as the latter controls the instant of \emph{nugget} formation.

If \cref{eq:nugget_condition} does not hold, the timescale of fermion density perturbation 
growth can be comparable to cosmological scales. As $M_\mathrm{eff}$
gets smaller, the long range interaction will generically compete with
gravity, leading to effects on LSS. In addition, as discussed in 
\cref{{sec:bkgequations}}, the background will behave like
quintessence and depend on the initial value of the scalar 
field. Both scenarios have been widely studied in the literature~\cite{Ratra:1987rm, Wetterich:1987fm, Frieman:1995pm, Coble:1996te, Caldwell:1997ii, Copeland:2006wr, Tsujikawa:2013fta, DAmico:2018hgc, Friedman:1991dj,Bean:2001ys,Gubser:2004uh,Nusser:2004qu,Bean:2008ac,Kesden:2006zb,Bai:2015vca,Mohapi:2015gua}.

\section{Cosmological Observables and Data Analysis: Neutrinos as a Benchmark}
\label{sec:analysis}
As discussed in the Introduction, we will now explore the observational consequences for neutrinos of the physics 
developed in \cref{sec:formalism}. From the results in that Section,
we conclude that neutrinos are particularly 
well suited to study the cosmology of long range interactions. On the one hand,
scalar long range interaction 
effects start being relevant when the fermion temperature drops below its mass. For neutrinos, this generically 
happens relatively late in the cosmological history, at times from which we have precise observations. On the other hand,
cosmology claims to be sensitive to the 
absolute neutrino mass scale, although cosmological observations do not kinematically measure the neutrino mass. As 
neutrinos decouple from the primordial plasma relatively early, CMB
and LSS observations are only sensitive to their gravitational 
impact. That is, their energy density and how it changes as the Universe expands (i.e., their equation of state). This 
depends on the neutrino mass in an assumption-dependent way~\cite{Oldengott:2019lke,Cuoco:2005qr}, and we expect this 
measurement to be affected by the modified equation of state induced by long range interactions (see 
\cref{fig:ew_nuggets}).

For simplicity, we will consider a single scalar field universally
coupled to all three neutrino mass eigenstates, which we assume to be
degenerate in mass with individual masses $m_0$. As 
discussed in \cref{sec:bkgequations}, we are interested in studying the regime where neutrino-neutrino scatterings can be 
neglected, and when the scalar mass $M_\mathrm{eff}$ is much larger
than the Hubble parameter. The former assumption corresponds to coupling constants 
${g \lesssim 10^{-7}}$~\cite{Hannestad:2005ex},
guaranteeing that the neutrino momentum distribution 
is the same as in the Standard Model. To a good approximation, this corresponds to a Fermi-Dirac distribution with 
negligible chemical potentials and a present day temperature $T_\nu \simeq 0.716 \, T_\gamma$~\cite{deSalas:2016ztq,Akita:2020szl,Froustey:2020mcq,Bennett:2020zkv,Escudero:2020dfa} with 
$T_\gamma$ the photon temperature. For $g \lesssim 10^{-7}$, $\frac{g m_0}{M_\phi} > 1$ and $m_0 \sim 0.1 \, \mathrm{eV}$;
$M_\mathrm{eff} \gg H$ at the relevant temperatures $T \lesssim m_0$ implies $10^{-8} \, \mathrm{eV} \gtrsim M_\phi \gtrsim 10^{-25} \, \mathrm{eV}$. These $\sim 17$ orders of magnitude in mediator mass have not been systematically studied in the literature and, as we will see, they can impact cosmological observations.

Regarding laboratory constraints, couplings $g \lesssim 10^{-7}$ are
well allowed~\cite{Lessa:2007up,Blinov:2019gcj,Pasquini:2015fjv,Agostini:2015nwa,Blum:2018ljv,Brune:2018sab,Escudero:2019gfk,Brdar:2020nbj}. Nevertheless, for small $M_\phi$ long range interactions may affect neutrino oscillation data. On the one hand, the 
cosmic neutrino background could reduce the present day effective neutrino mass below the minimum value allowed by 
oscillations. On the other hand, the large neutrino number density in the Sun could reduce the effective neutrino 
mass, modifying solar neutrino
data~\cite{Barger:2005mn,Cirelli:2005sg,Babu:2019iml}.\footnote{This effect can be more important in supernovae due to 
larger neutrino densities. Current SN1987A data is compatible with
massless neutrinos, and thus insensitive to these long range effects, but future observations might be 
sensitive to them.} Both effects are relevant only for $\frac{g}{M_\phi} 
\gtrsim 10^{5}$--$10^6 \, \mathrm{eV}^{-1}$ and, furthermore, they can be easily avoided by modifying the flavor structure of 
the scalar-neutrino coupling. Therefore, we will mostly ignore them in
what follows.

In this Section, we will study the impact of neutrino long range
interactions on CMB anisotropies, Baryon Acoustic Oscillation (BAO) data, and future 
LSS observations. We will start by qualitatively understanding the physical effects. 
We will then perform a Bayesian analysis of the Planck 2018 TT, TE, EE, 
lowE, and lensing CMB data~\cite{Aghanim:2018eyx}; as well as the BAO data from the 
6dF galaxy survey~\cite{Beutler:2011hx}, the Main Galaxy Sample from the SDSS DR7~\cite{Ross:2014qpa}, and 
the BOSS-DR12 analysis~\cite{Alam:2016hwk}. Finally, we will study the prospects of adding data from the 
future Large Scale Structure (LSS) EUCLID
survey~\cite{Laureijs:2011gra,Sprenger:2018tdb,Audren:2012vy}. To
carry out these analyses, we 
have modified the publicly available \verb+CLASS+ 
code~\cite{Lesgourgues:2011re,Blas:2011rf,Lesgourgues:2011rg,Lesgourgues:2011rh} to solve the cosmological 
perturbation equations with long range interactions (our modification is available \href{https://github.com/jsalvado/class_public_lrs}{at this URL} \github{jsalvado/class_public_lrs}), 
and we have explored the parameter space with the public Markov Chain Monte Carlo (MCMC) code \verb+Monte Python+~\cite{Audren:2012wb,Brinckmann:2018cvx}. All MCMC chains have been run until every Gelman-Rubin 
coefficient~\cite{Gelman:1992zz} was $R-1 < 0.02$. Our priors on the model parameters are summarized in 
\cref{tab:priors}. In particular, the range of $\sum m_\nu$ covers all values allowed by 
oscillations~\cite{Esteban:2020cvm,deSalas:2020pgw,Capozzi:2017ipn} and the latest results from the KATRIN experiment~\cite{Aker:2019uuj}. 
As we shall see, there are unbounded directions in the $\{\sum m_\nu,
g/M_\phi\}$ parameter space. To efficiently explore it, we have chosen
logarithmic priors in these parameters, as well as parameter ranges
that avoid excessive Bayesian volume effects.

\begin{table}[hbtp]
    \centering
    \begin{tabular}{cccl}
    \toprule
    Parameter & Prior & Range & Meaning\\
    \midrule
    $\omega_\mathrm{b}$ & Linear & $[0, \infty)$ & Reduced baryon density parameter\\
    $\omega_\mathrm{cdm}$ & Linear & $[0, \infty)$ & Reduced cold dark matter density parameter\\
    $\theta_s$ & Linear & $[0, \infty)$ & Acoustic CMB angular scale\\
    $A_s$ & Logarithmic & $[0, \infty)$ & \begin{tabular}[l]{@{}l@{}}Primordial power spectrum amplitude at \\comoving scale $k_0 = 0.05 \, \mathrm{Mpc}^{-1}$\end{tabular} \\
    $n_s$ & Linear & $[0, \infty)$ & Scalar spectral index\\
    $\tau_\mathrm{reio}$ & Linear & $[0.004, \infty)$ & Optical depth to reionization\\
    $\sum m_\nu / \mathrm{eV}$ & Logarithmic & $[0.024, 3]$ & Sum of neutrino masses\\
    $\frac{g}{M_\phi} \times \mathrm{eV}$ & Logarithmic & $[10^{-2},
                  10^{7.5}]$ & \begin{tabular}[l]{@{}l@{}}Long range interaction coupling divided
                by \\ the mediator mass\end{tabular}\\
    \bottomrule
    \end{tabular}
    \caption{Model parameters in our analysis along with their priors,
      ranges and physical meanings. For a parameter $x$, ``Linear'' prior means that we take a uniform prior on $x$, whereas for ``Logarithmic'' we take a uniform prior on $\log x$.}
    \label{tab:priors}
\end{table}

\subsection{Analysis of Present Data}
\label{sec:data_now}

We begin by qualitatively understanding the main effects of neutrino long range interactions on CMB anisotropies. To this 
purpose, we show in \cref{fig:CMB} the CMB temperature power spectrum for $\Lambda$CDM with massless neutrinos (top), 
as well as its relative difference with respect to a model with long range interacting massive neutrinos 
with different interaction strengths (bottom). In order to mimic the observable effects, in the bottom 
panel we have kept fixed the well-measured parameters $\omega_b$,
$\omega_\mathrm{cdm}$, $\theta_s$, $A_s$, $n_s$, and
$\tau_\mathrm{reio}$ (see \cref{tab:priors} for the meaning of each
parameter). We also show in grey Planck 2018
data~\cite{Aghanim:2018eyx}. We start by reviewing the main effects of neutrino masses~\cite{Lesgourgues:2018ncw} (dashed orange line in 
\cref{fig:CMB}):

\begin{figure}[hbtp]
    \centering
    \includegraphics[width=\textwidth]{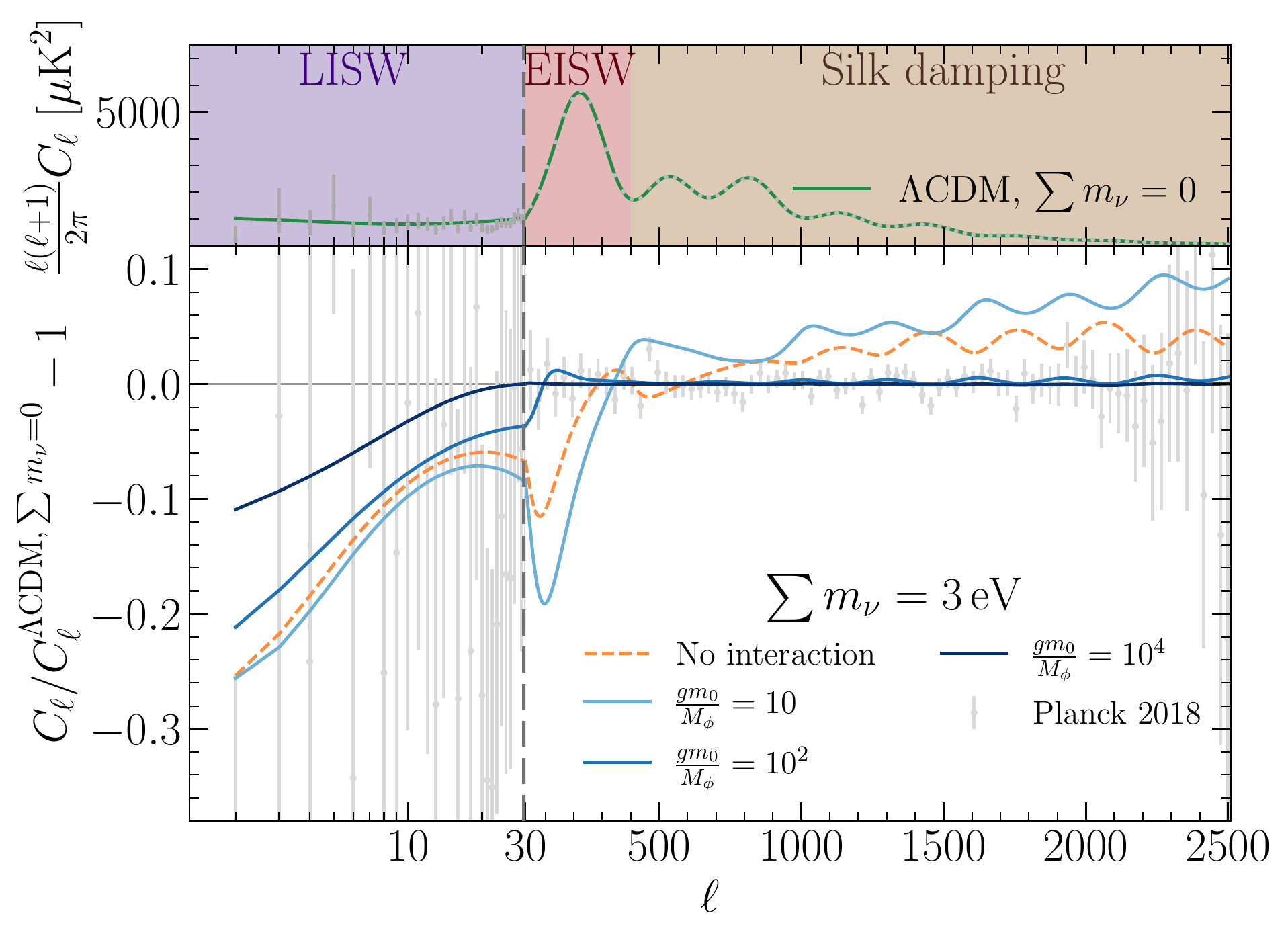}
    \caption{Top: CMB temperature power spectrum for $\Lambda$CDM with massless neutrinos [solid] and Planck 2018 data~\cite{Aghanim:2018eyx}. We show in shaded the multipole $\ell$ ranges where the Late Integrated Sachs-Wolfe effect (LISW), Early Integrated Sachs-Wolfe effect (EISW), and Silk damping leave their main imprints. Bottom: relative difference of the power spectrum between $\Lambda$CDM with massless neutrinos, and massive neutrinos with long range interactions. For the latter, we have chosen $\sum m_\nu = 3 \, \mathrm{eV}$ and different interaction strengths [solid blue shades]. We also show the result without interactions [dashed orange] as well as Planck 2018 data [grey].}
    \label{fig:CMB}
\end{figure}

\begin{itemize}
    \item \emph{The Integrated Sachs-Wolfe (ISW) effect}. This arises
      because, after leaving the last scattering surface, CMB photons
      traverse gravitational potential wells. Because of gravitational
      growth and the expansion of the Universe, the depth of the wells
      may change while photons are inside them. Therefore, the net
      gravitational redshift of photons after entering and exiting
      the wells may be non zero.
      
    This effect depends on the expansion rate of the Universe, i.e.,
    on its equation of state. It exactly vanishes for a fully matter
    dominated Universe, and is generically non-zero at two times: just
    after recombination, when the Universe still contains a
    non-negligible amount of radiation (Early Integrated Sachs-Wolfe,
    or EISW, effect); and at late times when the cosmological constant
    $\Lambda$ starts being relevant (Late Integrated Sachs-Wolfe, or
    LISW, effect).

    The contribution of neutrino masses to the LISW effect can be understood as follows. When neutrinos become non-relativistic, their energy density redshifts slower as their equation of state changes from radiation ($w=1/3$) to dust ($w=0$). Therefore, they will contribute more to the expansion of the Universe. This would modify the well-measured angular scale of the CMB peaks, $\theta_s = \left[\int_{z_\mathrm{rec}}^\infty c_s(z) \frac{\mathrm{d}z}{H(z)}\right]  \times \left[\int_0^{z_\mathrm{rec}} \frac{\mathrm{d}z}{H(z)}\right]^{-1}$ with $z_\mathrm{rec}$ the recombination redshift and $c_s$ the speed of sound of the primordial plasma, and can be compensated for by changing $\Lambda$ and therefore modifying the LISW effect. This is visible in the low $\ell$ region of \cref{fig:CMB}.
    
    Similarly, the EISW effect measures how the equation of state of the Universe deviates from $w=0$ close to recombination. If neutrinos become non-relativistic early enough, their contribution to the EISW effect will be reduced: this is visible for $\ell \sim 200$ in \cref{fig:CMB}.
    
    \item \emph{Silk damping}. Due to the non-zero photon mean free
      path, perturbations at small angular scales (large $\ell$) are
      exponentially damped. The characteristic angular scale of this
      damping, $\theta_D \propto \sqrt{\int_{z_\mathrm{rec}}^\infty
        \frac{1}{a(z) n_e(z)} \frac{\mathrm{d}z}{H(z)}}$ with $n_e$
      the free electron number density, depends on the neutrino
      contribution to the expansion of the Universe before
      recombination. Thus, if neutrinos become non-relativistic before recombination, their energy density redshifts slower, they contribute more to $H(z)$, and the damping scale gets reduced. This is visible at large $\ell$ in \cref{fig:CMB}.
\end{itemize}

We therefore conclude that the CMB measurement of neutrino masses is mostly a measurement of their equation 
of state as a function of redshift. Thus, the non-trivial equation of state that our model introduces (see 
\cref{fig:ew_nuggets}) will affect the same three CMB features discussed above. This is visible in 
\cref{fig:CMB}: for $\frac{g m_0}{M_\phi}=10$ neutrinos behave as dust earlier, enhancing the effects of 
neutrino masses; for $\frac{g m_0}{M_\phi}=10^2$ there is a period where $w<0$ and therefore the EISW 
effect is enhanced, contrarily to the effect of massive neutrinos; and for $\frac{g m_0}{M_\phi}=10^4$ the 
neutrino system behaves as radiation before recombination, removing the EISW and Silk damping effects of 
neutrino masses. We anticipate from these results that large interaction strengths will significantly 
affect the cosmological bound on neutrino masses, as they delay and modify the $w=1/3$ to 
$w=0$ transition. A similar effect was explored in Refs.~\cite{Oldengott:2019lke,Cuoco:2005qr}, where this transition changed due 
to a non-thermal neutrino distribution function.

\begin{figure}[hbtp]
    \centering
    \includegraphics[width=\textwidth]{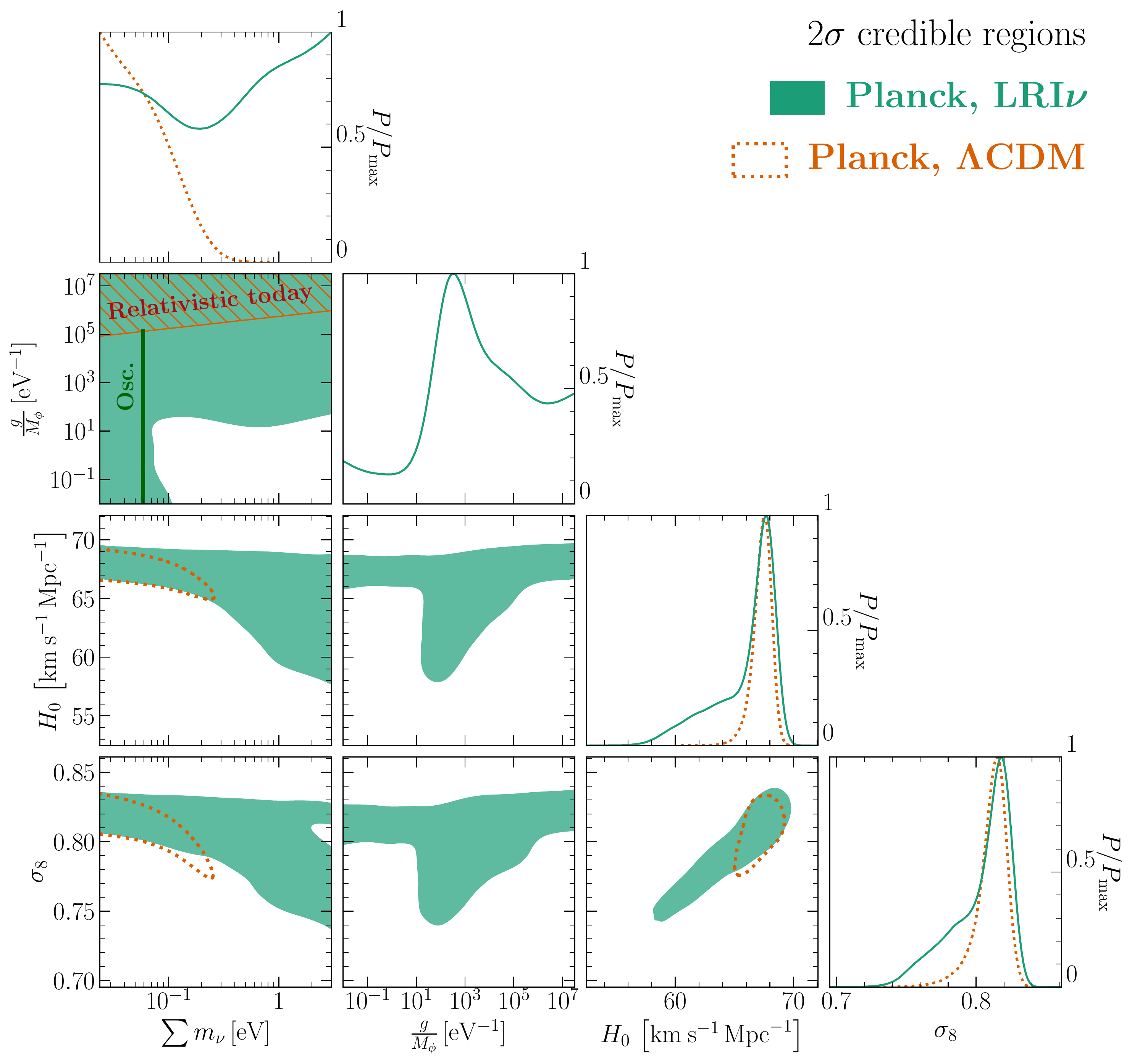}
    \caption{Planck 2018 constraints~\cite{Aghanim:2018eyx} on long
      range interacting neutrinos, LRI$\nu$, [solid light green] and on $\Lambda$CDM with non-interacting massive neutrinos [dotted orange]. We show the marginalized $2\sigma$ credible regions and 1-D posterior probability distributions for the most relevant parameters. For $\sum m_\nu$ and $\frac{g}{M_\phi}$, posteriors are constructed with uniform logarithmic bins. In the hatched region, cosmic neutrinos are relativistic today. The dark green line is the minimum value of $\sum m_\nu$ allowed by neutrino oscillation data~\cite{Esteban:2020cvm,deSalas:2020pgw,Capozzi:2017ipn}.}
    \label{fig:triangle_planck}
\end{figure}

Moving on to the data analysis, we show in \cref{fig:triangle_planck} the results of analyzing Planck CMB observations. In solid, we show
the 1-D posterior probabilities and the marginalized 2-D $2\sigma$ credible regions for the parameters that 
are most affected by our modification to $\Lambda$CDM: the sum of
neutrino masses $\sum m_\nu$, the interaction 
strength as parametrized by $\frac{g}{M_\phi}$, the Hubble constant $H_0$, and the amplitude parameter 
$\sigma_8$. In dotted, we show the results assuming $\Lambda$CDM with massive neutrinos; and in the hatched 
region cosmic neutrinos would still be relativistic today. As discussed before, a priori the hatched region 
is in conflict with neutrino oscillation measurements, although simple modifications of the coupling 
structure could accommodate this data. The dark green line is the minimum value of $\sum m_\nu$ allowed by neutrino oscillation data~\cite{Esteban:2020cvm,deSalas:2020pgw,Capozzi:2017ipn}. For completeness, we show in 
\cref{sec:app_triangles} the posterior probabilities and credible regions for all parameters in our 
analysis.

As we see, no neutrino mass bound can be obtained from CMB data if the interaction is strong enough to delay
the relativistic to non-relativistic equation of state transition (see \cref{fig:backgroundB}): for $g/M_\phi \gtrsim 10^2 \, 
\mathrm{eV}^{-1}$, the neutrino system still behaves as radiation at recombination. For small couplings $g/M_\phi 
\lesssim 10 \, \mathrm{eV}^{-1}$, we essentially recover the standard cosmology and neutrino mass bound. 
Because of this, for $\sum m_\nu \gtrsim 0.1 \, \mathrm{eV}$ interaction strengths $g/M_\phi \lesssim 10 \, 
\mathrm{eV}^{-1}$ are excluded.

We also observe that $H_0$ and $\sigma_8$ are quite correlated with the neutrino mass and interaction strength. 
The correlation with $\sum m_\nu$ is also present in the standard $\Lambda$CDM scenario [dotted orange], 
and is due to the late time contribution of massive neutrinos to the
energy density of the Universe. The correlation
with $g/M_\phi$, in turn, gets reduced at large couplings. This is because for
such couplings our attractive self 
interaction reduces the energy density in neutrinos, partly due to \emph{nugget} formation (see 
\cref{fig:ew_nuggets}).

The allowed parameter values in \cref{fig:triangle_planck} have essentially the same cosmological history 
before recombination: neutrinos behaving as radiation. As they differ in their post-recombination behavior, 
we expect late time cosmological probes to be sensitive to a region of parameter space allowed by Planck 
data. 

{
LSS measurements are a standard but powerful example of such probes. They contain many rich features, sensitive both to the late-time structure growth and to the expansion history of the Universe. Furthermore, there are very precise present observations available that will significantly improve in the near future~\cite{Beutler:2011hx,Ross:2014qpa,Alam:2016hwk,Abbott:2017wau,Troxel:2017xyo,Asgari:2020wuj,Hikage:2018qbn,Laureijs:2011gra,Maartens:2015mra,Mandelbaum:2018ouv}. As the goal of this work is not to carry out a detailed study of the complementarity among different datasets, we will only use BAO results for our analysis of present data. These are accurate and generically accepted to be robust against systematic uncertainties and changes in the underlying cosmological model. Including other LSS probes may require a more careful treatment of the data and the systematic uncertainties. Furthermore, our scenario and non-interacting massive neutrinos induce similar features in the matter power spectrum. Including additional LSS data does not significantly improve the neutrino mass bound~\cite{Aghanim:2018eyx}, partially due to small tensions with Planck, and so we don't expect it to change our conclusions regarding neutrino long range interactions.}

\begin{figure}[hbtp]
    \centering
    \includegraphics[width=\textwidth]{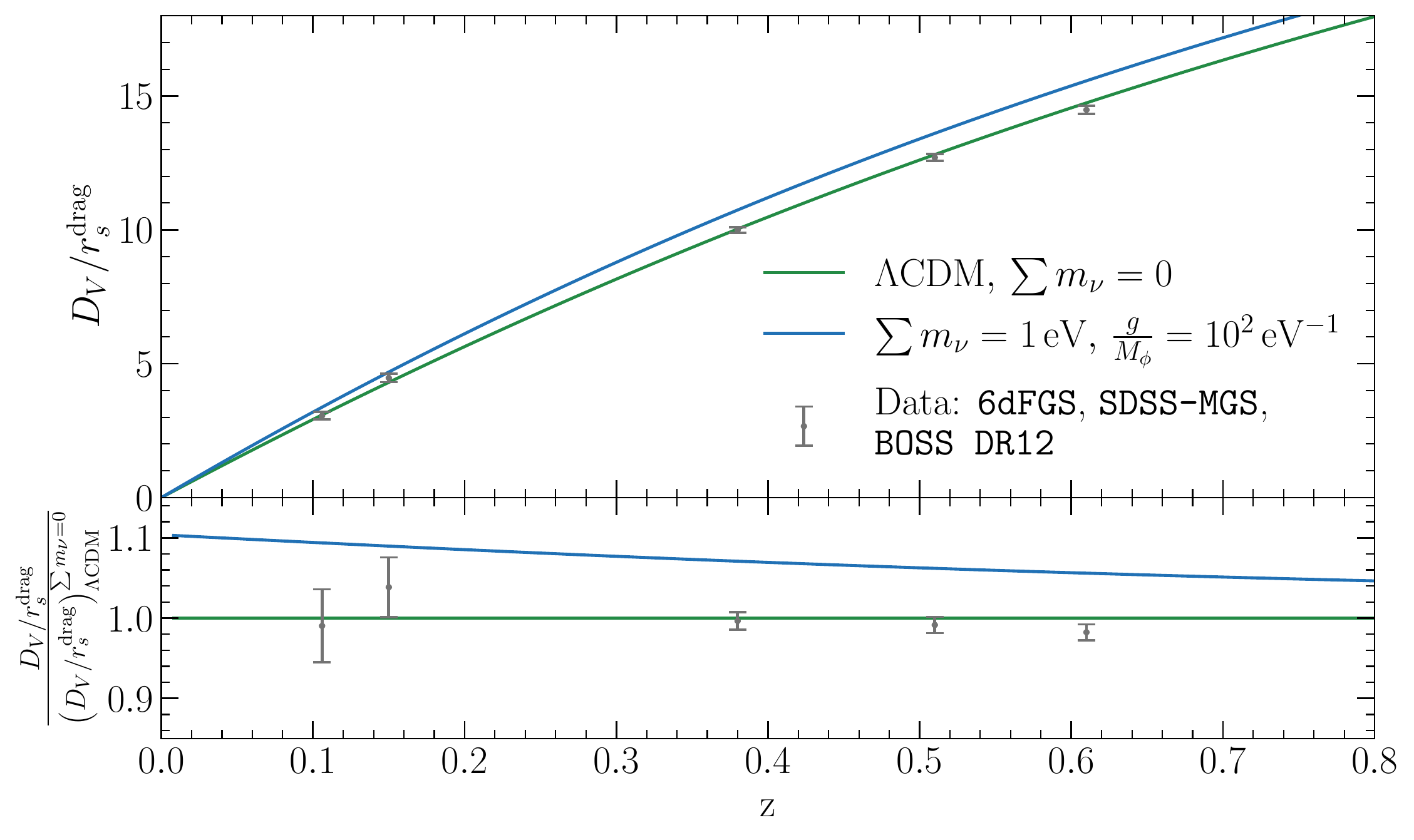}
    \caption{Top: BAO scale for $\Lambda$CDM with massless neutrinos [green], with a model of self-interacting massive neutrinos allowed by Planck data [blue], and observational data~\cite{Beutler:2011hx,Ross:2014qpa,Alam:2016hwk}. Bottom: ratios with respect to $\Lambda$CDM with massless neutrinos. To generate the solid curves, we have fixed $\omega_b$, $\omega_\mathrm{cdm}$, $\theta_s$, $A_s$, $n_s$, and $\tau_\mathrm{reio}$.}
    \label{fig:BAO}
\end{figure}

BAO measurements are mostly sensitive to 
\begin{equation}
\frac{D_V}{r_s^\mathrm{drag}} (z) = \frac{\displaystyle\left[\frac{z}{H(z)} \left( \int_0^z \frac{\mathrm{d}z'}{H(z')} \right)^2\right]^{1/3}}{\displaystyle\int_{z_\mathrm{drag}}^\infty c_s(z') \frac{\mathrm{d}z'}{H(z')}} \, ,
\label{eq:BAO_scale}
\end{equation}
with $z_\mathrm{drag}$ the baryon drag redshift. We show in \cref{fig:BAO} this quantity as a function of 
redshift for two scenarios: $\Lambda$CDM with massless neutrinos, and $\Lambda$CDM with massive 
self-interacting neutrinos. For the latter, we have chosen $\sum m_\nu = 1 \, \mathrm{eV}$ and 
$\frac{g}{M_\phi} = 10^2 \, \mathrm{eV}^{-1}$, parameter values allowed by Planck data (see 
\cref{fig:triangle_planck}) for which neutrinos behave as radiation before recombination. The difference 
between both curves is therefore due to the neutrino contribution to the late time energy density of the 
Universe, and thus to the late time Hubble parameter in \cref{eq:BAO_scale}. We also show observational 
data from Refs.~\cite{Beutler:2011hx,Ross:2014qpa,Alam:2016hwk}, in clear tension with the self-interacting 
neutrino scenario.

\begin{figure}[hbtp]
    \centering
    \includegraphics[width=\textwidth]{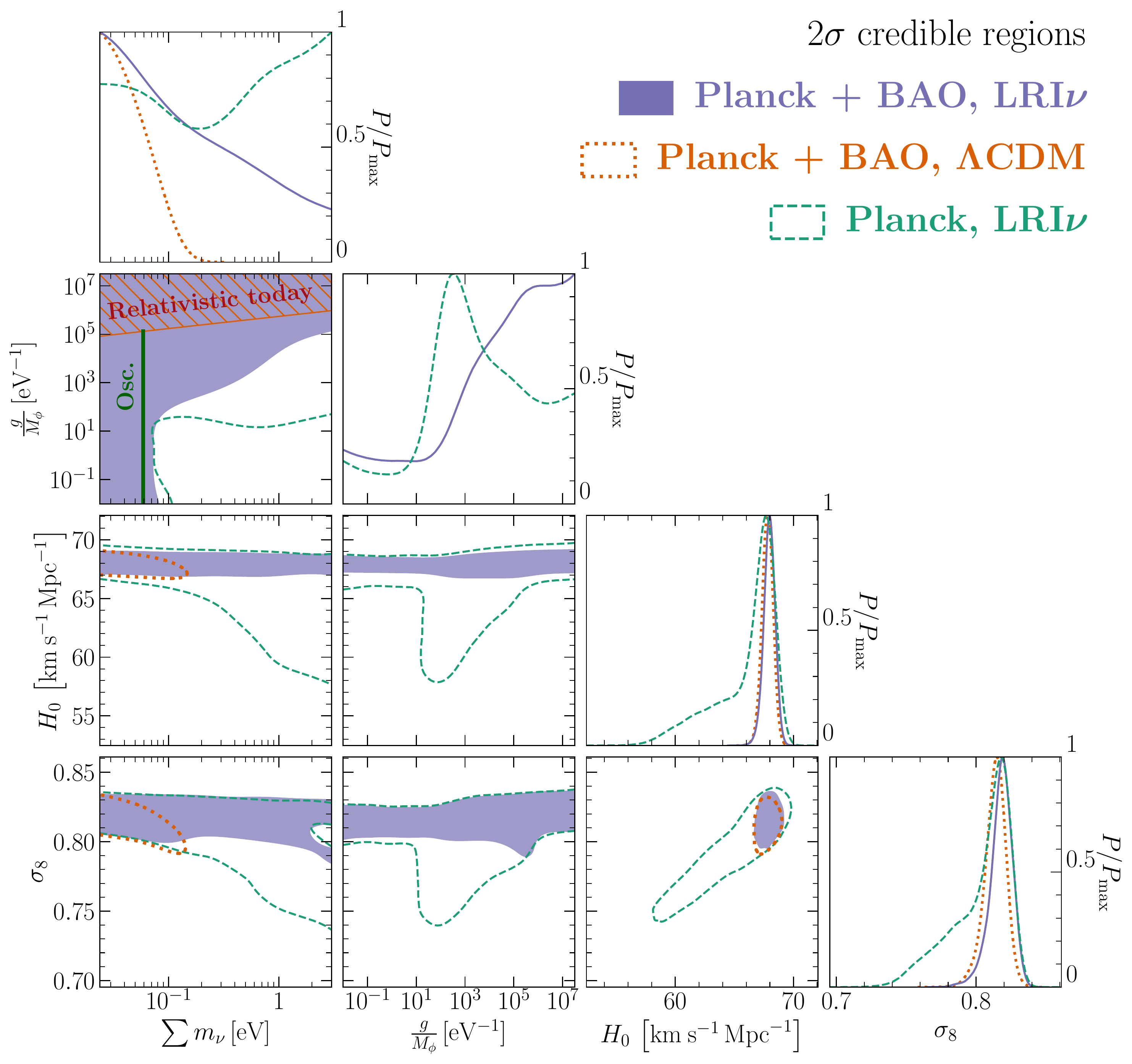}
    \caption{Planck 2018+BAO constraints~\cite{Aghanim:2018eyx,Beutler:2011hx,Ross:2014qpa,Alam:2016hwk} on 
      long range interacting neutrinos, LRI$\nu$, [solid purple], on $\Lambda$CDM with non-interacting massive neutrinos 
    [dotted orange], and Planck 2018
    constraints~\cite{Aghanim:2018eyx} on long range interacting neutrinos 
    [dashed green]. We show the marginalized $2\sigma$ credible regions and 1-D posterior probability 
    distributions for the most relevant parameters. For $\sum m_\nu$ and $\frac{g}{M_\phi}$, posteriors are 
    constructed with uniform logarithmic bins. In the hatched region, cosmic neutrinos are relativistic 
    today. The solid dark green line is the minimum value of $\sum m_\nu$ allowed by neutrino oscillation data~\cite{Esteban:2020cvm,deSalas:2020pgw,Capozzi:2017ipn}.}
    \label{fig:triangle_planck_BAO}
\end{figure}

Regarding the full data analysis, we show in \cref{fig:triangle_planck_BAO} the 1-D posterior probabilities and the 
marginalized 2-D $2\sigma$ credible regions including Planck 2018 and BAO 
observations~\cite{Aghanim:2018eyx,Beutler:2011hx,Ross:2014qpa,Alam:2016hwk} for $\sum m_\nu$, $\frac{g}{M_\phi}$, 
$H_0$, and $\sigma_8$. We show in solid the results assuming $\Lambda$CDM with massive self-interacting 
neutrinos, in dotted for $\Lambda$CDM with massive neutrinos without self interactions, in dashed for 
$\Lambda$CDM with massive self-interacting neutrinos but without BAO 
data (i.e., the light green lines in \cref{fig:triangle_planck}), and in the hatched region cosmic neutrinos 
would still be relativistic today. The dark green line is the minimum value of $\sum m_\nu$ allowed by neutrino oscillation data~\cite{Esteban:2020cvm,deSalas:2020pgw,Capozzi:2017ipn}. For completeness, we show in \cref{sec:app_triangles} the posterior 
probabilities and credible regions for all parameters in our analysis. 

As we see, BAO data excludes a large amount of interaction strengths for relatively large values of the 
neutrino mass. As discussed above, this is due to the dependence of the late time neutrino energy density 
on $\sum m_\nu$ and $g/M_\phi$, partly due to \emph{nugget} formation (see \cref{fig:ew_nuggets}). 
Excluding a large amount of interaction strengths also breaks the degeneracies with $H_0$ and $\sigma_8$, 
giving essentially the same results as $\Lambda$CDM for these parameters. Nevertheless, there is still 
no cosmological neutrino mass bound.

In other words, the KATRIN laboratory experiment, that aims to constraint 
$\sum m_\nu \lesssim 0.6 \, \mathrm{eV}$~\cite{Aker:2019uuj}, could in
the near future detect a non-zero neutrino mass 
compatible with cosmology for interaction strengths $g/M_\phi \sim 10^3\textup{--}10^6 \, \mathrm{eV}^{-1}$.

Notice that, especially after introducing BAO data, long range interacting neutrinos do not solve the $H_0$ 
tension (see Ref.~\cite{Verde:2019ivm} for an overview of the tension and proposed solutions).\footnote{ As can be seen from \cref{fig:triangle_planck_BAO_full}, the posteriors for $\Omega_m$ and $\sigma_8$ are essentially the same as in $\Lambda$CDM. Thus, we expect the parameter $S_8 \equiv \sigma_8 \sqrt{\Omega_m/0.3}$ and its associated tension~\cite{Aghanim:2018eyx} not to be affected by neutrino long range interactions.} This can be understood
from \cref{fig:ew_nuggets}: this tension is generically solved by increasing the energy density of the Universe around
recombination, but a scalar interaction, being 
universally attractive, will \emph{reduce} the energy.

\subsection{Future Prospects for Large Scale Structure}
\label{sec:data_future}

As we have just discussed, since neutrinos become non-relativistic relatively late, late time cosmological 
measurements are generically  quite sensitive to neutrino long range interactions. In the previous subsection, we 
have illustrated this point with BAO data, a precise feature of the matter power spectrum that is considered 
to be robust. The situation should further improve in the near future, when surveys such as EUCLID~\cite{Laureijs:2011gra} aim to 
precisely measure the full power spectrum at different redshifts. In the following, we will study the impact of long range interacting neutrinos on 
the matter power spectrum, as well as the implications of future EUCLID data.

\begin{figure}[hbtp]
    \centering
    \includegraphics[width=0.85\textwidth]{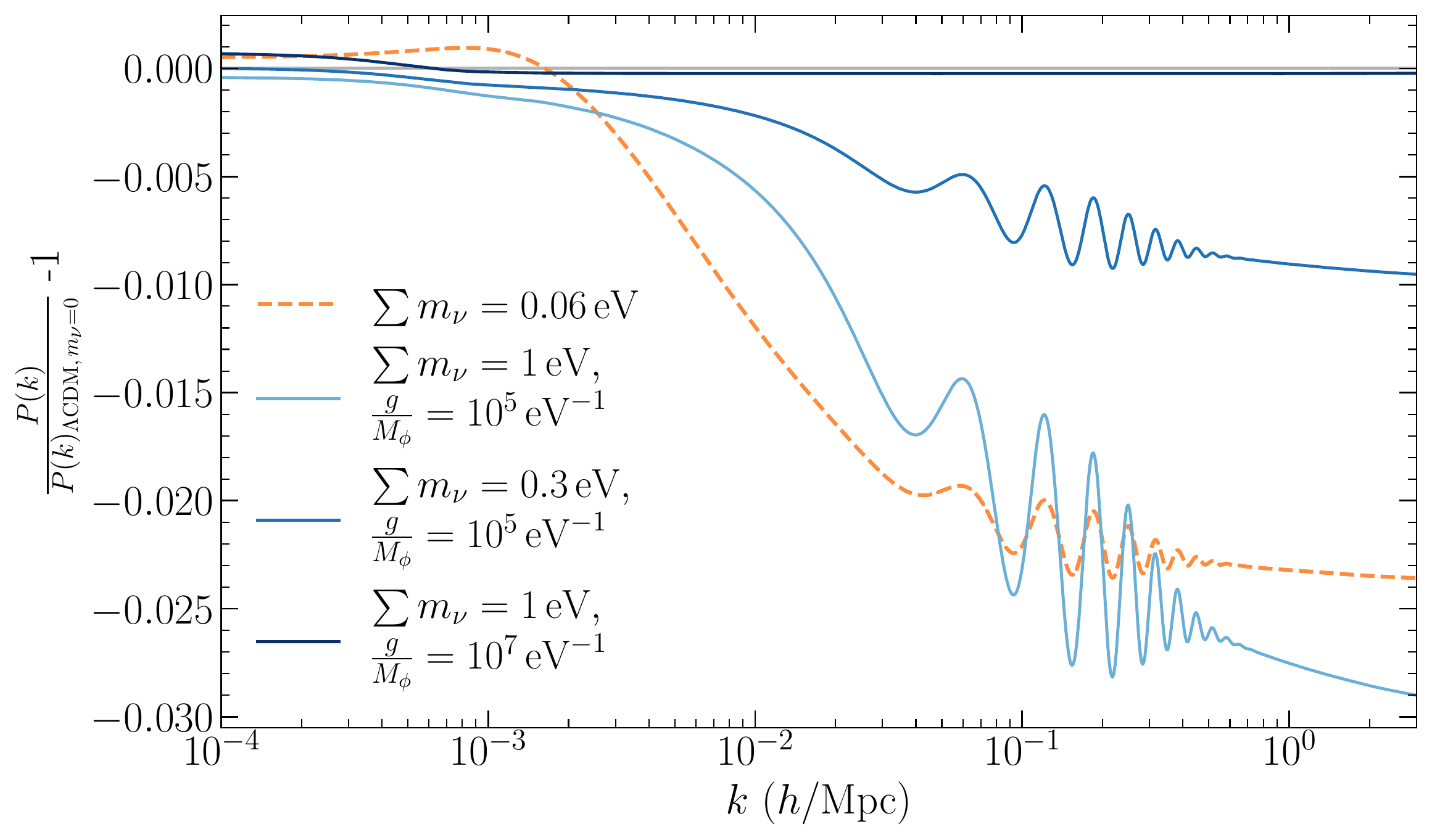}
    \caption{Relative difference in linear matter power spectrum between $\Lambda$CDM with massless neutrinos and massive 
    neutrinos with total mass $\sum m_\nu$ and interaction long range strength $\frac{g}{M_\phi}$. The dashed orange line 
    corresponds to non-interacting massive neutrinos with the smallest mass allowed by oscillations, which should 
    be detectable at $\sim 2\sigma$--$3\sigma$ with EUCLID~\cite{Hamann:2012fe,Sprenger:2018tdb}. To generate the curves, we have fixed 
    $\omega_b$, $\omega_\mathrm{cdm}$, $H_0$, $A_s$, $n_s$, and $\tau_\mathrm{reio}$. The non-zero values at low 
    $k$ are within numerical uncertainties.}
    \label{fig:LSS}
\end{figure}

We start by illustrating in \cref{fig:LSS} the relative difference in matter power spectrum between $\Lambda$CDM with 
massless neutrinos, and $\Lambda$CDM with self-interacting massive neutrinos for different masses and long range interaction strengths. In dashed 
orange, neutrinos do not self interact and have the smallest mass allowed by oscillation data. The solid blue 
lines correspond to different parameters allowed by Planck and BAO data (see \cref{fig:triangle_planck_BAO}).

For non-interacting massive neutrinos, we observe the well-known
enhancement at $k \sim 10^{-3} h/\mathrm{Mpc}$ and the suppression at
large $k$. The former is due to non-relativistic neutrinos falling in
the dark matter gravitational wells  
and thus contributing to structure growth. The latter is due to the massive neutrino contribution to the energy 
density of the Universe: massive neutrinos increase the Hubble parameter with respect to the massless scenario (as 
they have an equation of state $w<1/3$ and thus redshift slower), which in turn suppresses structure growth. Long range interactions 
delay the relativistic to non-relativistic transition (see \cref{fig:backgroundB}) and modify the equation of 
state (see \cref{fig:ew_nuggets}). Therefore, they remove the power spectrum enhancement at $k \sim 10^{-3} 
h/\mathrm{Mpc}$ and modify the Hubble friction-induced large $k$
suppression. Both effects are visible in the solid blue lines in 
\cref{fig:LSS}. The future EUCLID survey should have $\sim 2$--$3\sigma$ sensitivity to the dashed orange line~\cite{Hamann:2012fe,Sprenger:2018tdb},\footnote{ Present-day uncertainties on the matter power spectrum are larger than the range in \cref{fig:LSS}.} and so it could probe the parameters corresponding to the light blue line, allowed by current Planck and BAO data.

To quantitatively explore the potential of EUCLID, we have carried out a Bayesian analysis combining current Planck CMB data with 
an EUCLID power spectrum and lensing forecast following the prescriptions of 
Refs.~\cite{Sprenger:2018tdb,Audren:2012vy}. { We have included non-linearities using the \verb+HALOFIT+ semi-analytic prescription~\cite{Takahashi:2012em,Bird:2011rb}, but we have not added any theoretical errors associated to them}. Therefore, we have chosen a  minimum 
comoving scale $k_\mathrm{max} = 0.2 \, h/\mathrm{Mpc}$ ($0.5 \,h/\mathrm{Mpc}$) for the power spectrum (lensing) data. { This roughly corresponds to the scale at which present-day theoretical errors match the smallest EUCLID observational uncertainties~\cite{Sprenger:2018tdb}, so we don't expect the future data sensitivity to be very different from our results}. Finally, following Ref.~\cite{Sprenger:2018tdb} (see also Refs.~\cite{Castorina:2013wga,Castorina:2015bma,Vagnozzi:2018pwo,Raccanelli:2017kht}), we have only included the baryon and 
cold dark matter power spectrum in the EUCLID galaxy power spectrum determination.

\begin{figure}[hbtp]
\centering
\makebox[\linewidth][c]{%
\begin{subfigure}{0.6\textwidth}
    \centering
    \includegraphics[width=\textwidth]{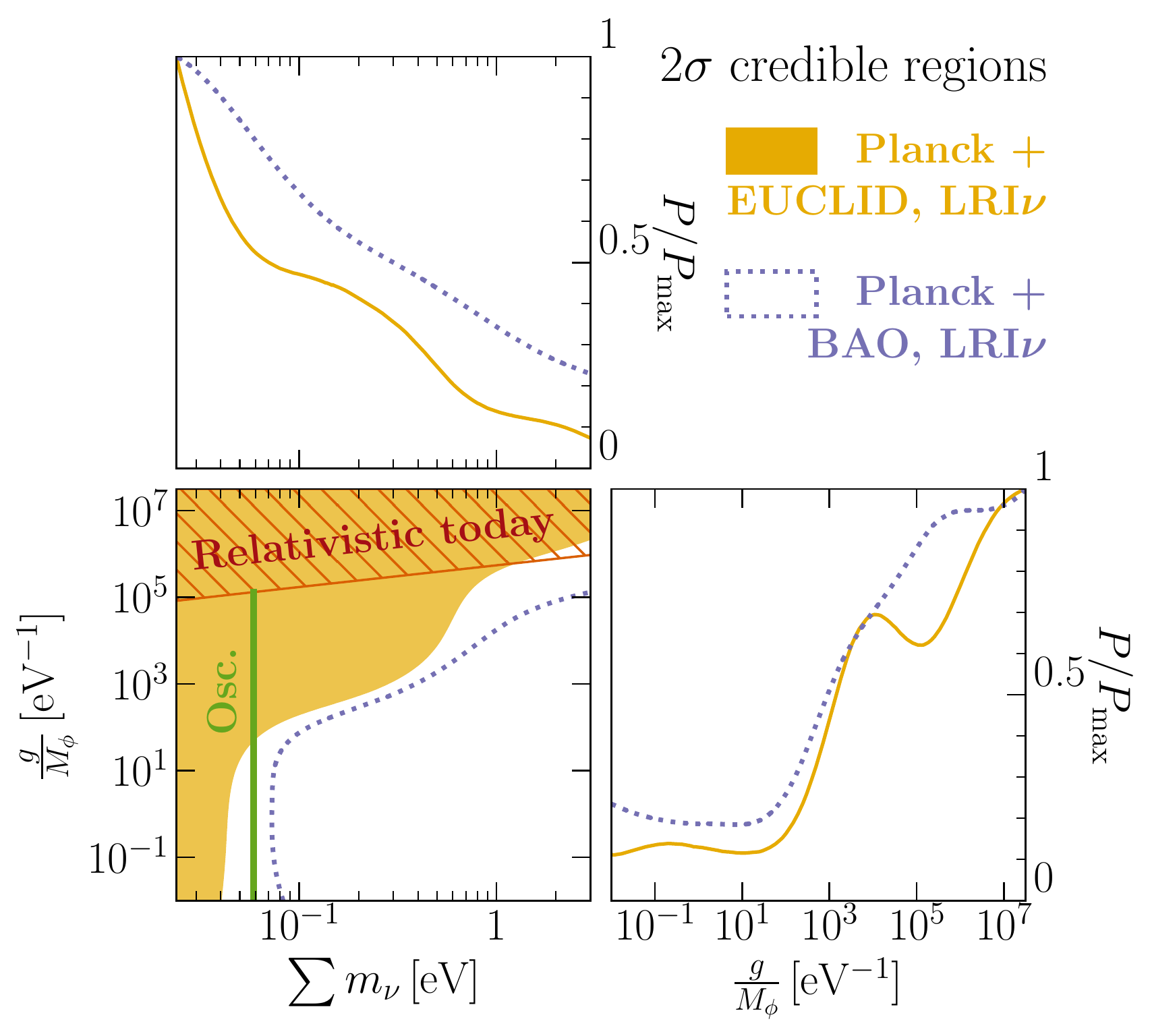}  
    \caption{$\sum m_\nu^\mathrm{true} = 0.024 \, \mathrm{eV}$}
    \label{fig:EuclidTrianglemassless}
\end{subfigure} \begin{subfigure}{0.6\textwidth}
    \centering
    \includegraphics[width=\textwidth]{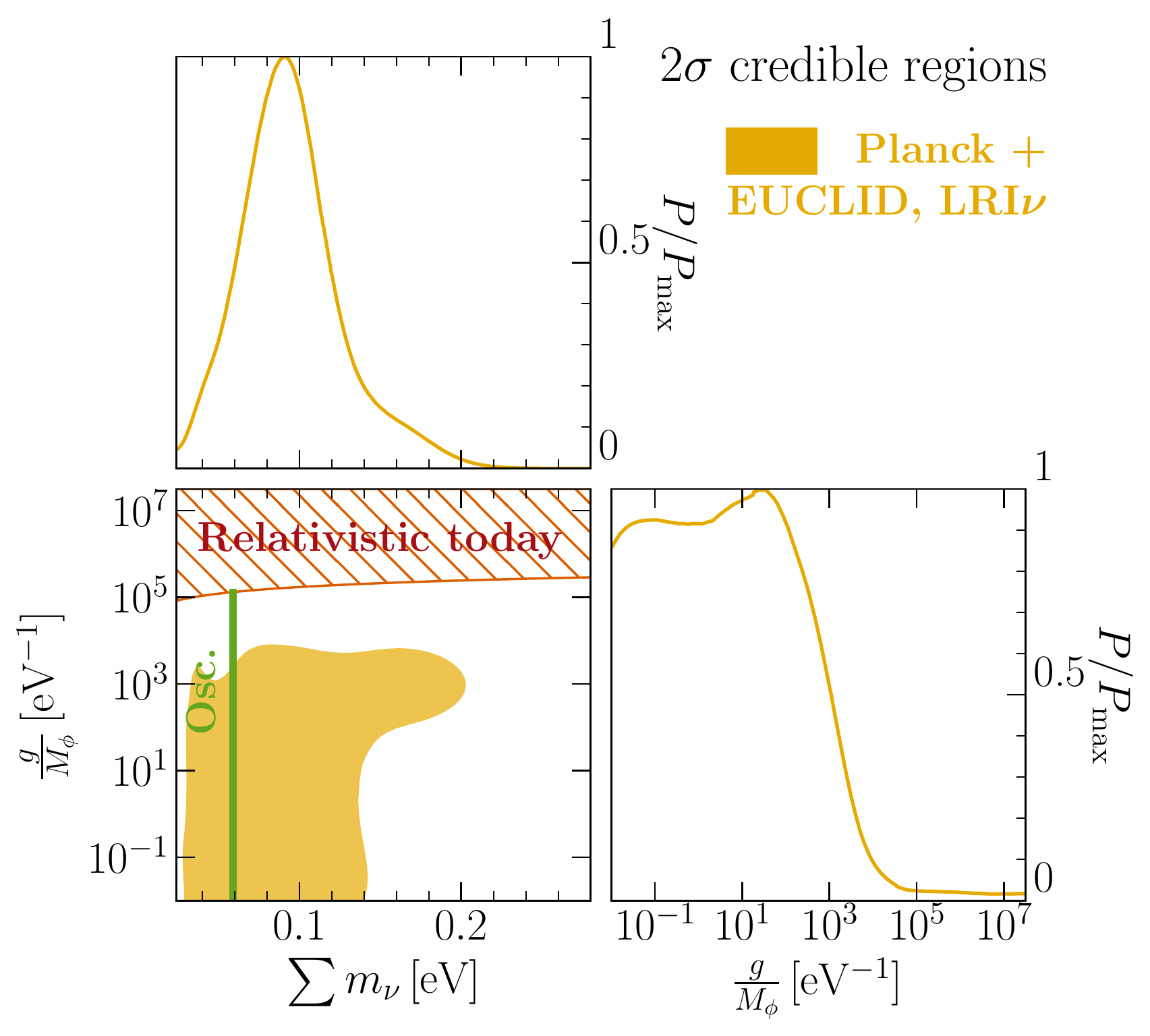}  
    \caption{$\sum m_\nu^\mathrm{true} = 0.08 \, \mathrm{eV}$}
    \label{fig:EuclidTrianglemassive}
\end{subfigure}
}
    \caption{Future EUCLID + Planck 2018 constraints~\cite{Laureijs:2011gra,Sprenger:2018tdb,Aghanim:2018eyx} on 
      long range interacting neutrinos, LRI$\nu$, [solid yellow], and Planck 2018 + BAO 
    constraints~\cite{Aghanim:2018eyx,Beutler:2011hx,Ross:2014qpa,Alam:2016hwk} [dotted purple]. We show the 
    marginalized $2\sigma$ credible regions and 1-D posterior probability distributions for $\sum m_\nu$ and 
    $\frac{g}{M_\phi}$. In the left panel, posteriors are constructed with uniform logarithmic bins. In the right panel, only the posterior on $\frac{g}{M_\phi}$ is constructed with logarithmic bins, and the prior on $\sum m_\nu$ is taken to be linear. In the hatched region, cosmic 
    neutrinos are relativistic today; and the green line is the minimum total neutrino mass allowed by oscillation 
    data~\cite{Esteban:2020cvm,deSalas:2020pgw,Capozzi:2017ipn}. We have generated EUCLID mock data with the best fit parameters from the 
    $\Lambda$CDM Planck 2018 + BAO analysis (last column in Table 2 in Ref.~\cite{Aghanim:2018eyx}), no long range 
    interactions and two different values of the total neutrino mass $\sum m_\nu^\mathrm{true}$: 0.024 eV [left], 
    the smallest value compatible with our priors, and consistent with massless neutrinos within EUCLID precision; 
    and 0.08 eV [right], a value compatible with CMB and BAO data that should be well-measurable by EUCLID.}
    \label{fig:EuclidTriangle}
\end{figure}

We show the results of our analysis in \cref{fig:EuclidTriangle}. In solid, we show the 1-D posterior 
probabilities and the marginalized 2-D $2\sigma$ credible regions for the sum of neutrino masses $\sum m_\nu$ and 
the long range interaction strength $\frac{g}{M_\phi}$. In dotted, we show the results from our Planck 2018 + BAO 
analysis (see \cref{fig:triangle_planck_BAO}). In the hatched region cosmic neutrinos are relativistic today, and 
the green line is the minimum value of $\sum m_\nu$ allowed by neutrino oscillation data~\cite{Esteban:2020cvm,deSalas:2020pgw,Capozzi:2017ipn}. 
We have generated EUCLID mock data using the best fit cosmological parameters of the Planck 2018 + BAO 
$\Lambda$CDM analysis~\cite{Aghanim:2018eyx}, no long range interactions, and two values for the sum of neutrino 
masses as labeled by the captions. On the one hand, the mock data for the analysis results in \cref{fig:EuclidTrianglemassless} has 
been generated with the smallest neutrino mass allowed by our priors in \cref{tab:priors}. This value is in direct 
tension with neutrino oscillation measurements and corresponds to EUCLID results compatible with massless 
neutrinos. On the other hand, in \cref{fig:EuclidTrianglemassive}, we have generated the data with $\sum m_\nu = 0.08 \, 
\mathrm{eV}$. This value is compatible with present cosmological
bounds and is well within the EUCLID sensitivity. For 
completeness, we show in \cref{sec:app_triangles} the posterior probabilities and credible regions for all 
parameters in our analysis.

\Cref{fig:EuclidTriangle} shows that, as expected, EUCLID data will improve the Planck 2018 + BAO 
constraints. Depending on the outcome of the EUCLID observations, we can consider two qualitatively distinct 
scenarios:
\begin{itemize}
    \item \emph{EUCLID data is consistent with massless neutrinos}. With the projected sensitivity, this would be 
    a contradiction between cosmological observations and neutrino
    oscillation experiments, and thus a hint for new 
    physics. The resulting prospects for an analysis assuming long range interacting neutrinos are shown in 
    \cref{fig:EuclidTrianglemassless}. As we see, long range interactions could explain the apparent 
    cosmology-oscillations discrepancy for interaction strengths 
    ${g/M_\phi \sim 10^2\textup{--}10^5 \, \mathrm{eV}^{-1}}$.
    \item \emph{EUCLID data is consistent with massive non-interacting neutrinos}. This is the expected outcome, 
    compatible with no new physics. The observed shape of the power
    spectrum (see \cref{fig:LSS}) would exclude neutrino long range interaction strengths ${g/M_\phi \gtrsim 10^4 \, \mathrm{eV}^{-1}}$. 
    Furthermore, even within our model, the measurement of the neutrino mass would be relatively robust, the upper limit being relaxed by $\sim 40\%$ for ${g/M_\phi \sim 10^3 \, \mathrm{eV}^{-1}}$.
\end{itemize}
Moreover, as mentioned in the previous subsection, a neutrino mass
detection at KATRIN of $\sum m_\nu \gtrsim 0.6 \, \mathrm{eV}$ could 
point to long range interactions with strength $g/M_\phi \sim 10^3 \textup{--} 10^5 \, \mathrm{eV}^{-1}$. As we 
see in \cref{fig:EuclidTriangle}, these parameter values can be explored by EUCLID, allowing to test this 
hypothesis.

\section{Summary and Conclusions}
\label{sec:conclusions}

In this work, we have consistently explored the cosmology of fermions endowed with a scalar-mediated long range 
interaction. We have then applied the general formalism to the particular case of self-interacting neutrinos, for which 
we have performed an analysis of current and near future cosmological data.

We have started by obtaining the evolution equations in \cref{sec:formalism}. We have focused on long range effects on the energy density and equation of state of the 
fermion system, that directly impact the cosmological evolution. For this, we have studied the regime where the 
effective scalar mass $M_\mathrm{eff}$~\eqref{eq:scalarMeff} is much larger than the Hubble parameter and 
collisions among fermions can be neglected. Relaxing the first assumption would recover quintessence and modified 
gravity-like scenarios, whereas relaxing the second assumption would introduce additional particle creation, 
annihilation and momentum transfer processes. Both limits have been widely studied in the literature~\cite{Ratra:1987rm, Wetterich:1987fm, Frieman:1995pm, Coble:1996te, Caldwell:1997ii, Copeland:2006wr, Tsujikawa:2013fta, DAmico:2018hgc, Friedman:1991dj,Bean:2001ys,Gubser:2004uh,Nusser:2004qu,Bean:2008ac,Kesden:2006zb,Bai:2015vca,Mohapi:2015gua,Beacom:2004yd,Chacko:2003dt,Chacko:2004cz,Hannestad:2004qu,Hannestad:2005ex,Bell:2005dr,Friedland:2007vv,Sola:2019jek,Sola:2020lba}.

To solve the evolution of the system, we have chosen as our initial conditions a Fermi-Dirac momentum distribution 
for the fermions, as well as no initial scalar field except for the one sourced by the fermions. Our main results 
are \cref{fig:background,fig:ew_nuggets}, where we show the total energy density and equation of state of the 
system as well as the effective fermion mass as a function of the
fermion temperature $T$. We have found 
that long range effects turn on for $T$ below the vacuum fermion mass $m_0$, and they are relevant if 
$\frac{g m_0}{M_\phi} > 1$. In other words, if for $T<m_0$ the interparticle distance is smaller than the 
interaction range.

As the Universe expands and the fermion temperature decreases, we have
obtained that the fermion system first behaves as
radiation even for temperatures well below $m_0$, as the sourced scalar field reduces the effective fermion mass. 
Later on, the scalar field energy density takes over and the equation
of state parameter $w$ can take negative values.
Finally, when the system cools down and the interparticle distance gets larger than the interaction range, 
fermions become effectively non-relativistic. At this point, the long range interaction is generically much 
stronger than gravity and fermion perturbations collapse in very short timescales ${\sim M_\mathrm{eff}^{-1} \ll 
H^{-1}}$, forming non-linear structures or \emph{nuggets} with typical sizes $\lesssim M_\mathrm{eff}^{-1}$ outside 
which no scalar field is left. Supported by analytic estimations, we have chosen to model this collapse as an 
\emph{instantaneous} transition to a dust-like behavior.

In \cref{sec:analysis}, we have confronted these new interactions with
data, focusing on long range interactions among neutrinos. We have discussed the consequences on CMB, BAO and LSS observables (\cref{fig:CMB,fig:BAO,fig:LSS}), as 
well as their physical origin. For this, we have modified the \verb+CLASS+ code to include long range interacting 
fermions. Our modification is publicly available \href{https://github.com/jsalvado/class_public_lrs}{at this URL} \github{jsalvado/class_public_lrs}. 
We have also performed a Bayesian analysis to present Planck 2018 and BAO data 
(\cref{fig:triangle_planck,fig:triangle_planck_BAO}). We have obtained that the cosmological neutrino mass bound is 
completely removed once long range interactions are included, due to
the effective neutrino mass induced by the scalar field. 
Thus, in our simple modification the KATRIN experiment could detect neutrino masses $\sim 1 \, \mathrm{eV}$ as long as
the self interaction strength is $g/M_\phi \sim 10^3\textup{--}10^6 \, \mathrm{eV}^{-1}$.

We have also concluded that BAO data plays an important role in breaking degeneracies. This is mostly due to BAO being a 
late-time cosmological probe, as neutrinos become non-relativistic relatively late. Because of this, we expect next 
generation LSS data to efficiently explore long range neutrino self interactions. In \cref{sec:data_future}, we have carried out a forecast of the future
EUCLID survey (\cref{fig:EuclidTriangle}), that aims to be sensitive enough to detect the smallest neutrino mass allowed by oscillations. 
Nevertheless, if EUCLID observations are compatible with massless neutrinos, we have found that long range interactions 
could explain the apparent discrepancy with oscillation experiments. If, in turn, EUCLID results are compatible with 
massive, non-interacting neutrinos, the long range interaction strength would be constrained to be $g/M_\phi \lesssim 10^4 \, \mathrm{eV}^{-1}$. In this scenario, a positive mass measurement would be quite robust against the presence of long 
range interactions, though the upper limit on the neutrino mass would be relaxed for $g/M_\phi \sim 10^3 \, \mathrm{eV}^{-1}$. Finally, if KATRIN measures a non-zero neutrino
mass, EUCLID could test whether the apparent discrepancy between KATRIN and
CMB and BAO data is due to long range interactions. 

In summary, in this work we have seen that long range interactions can
dramatically alter the equation of state of cosmological systems. By
dropping the ideal gas assumption, interacting fermion systems might behave as
ultrarelativistic at relatively low temperatures or even as dark
energy. For the case of neutrinos, cosmological probes sensitively
explore this physics, at the same time affecting the neutrino
mass bound. This opens the possibility for a laboratory detection of
the neutrino mass scale in the near future.  

\acknowledgments
We would like to thank M.~C.~Gonzalez-Garcia for very helpful comments and discussions and a careful reading of the 
manuscript, Alessio Notari for early discussions, and John Beacom for comments. This work has been funded by the European ITN project H2020-MSCA-ITN-2019/860881-HIDDeN, the Spanish grants FPA2016-76005-C2-1-P, 
PID2019-108122GB-C32, PID2019-105614GB-C21. IE acknowledges support from the FPU  
program fellowship FPU15/03697, and warmly thanks CCAPP for their valuable support and hospitality during the final stages of this work.

\clearpage
\appendix
\section{Classical Limit of the Evolution Equations}
\label{sec:app_classical}

The evolution of the fermion and scalar field is dictated by the quantum evolution equations 
\cref{eq:scalarMasterEq,eq:diracEq}. In this Appendix, we will obtain the classical limit relevant for 
the cosmological scales we are interested in.

As discussed in \cref{sec:formalism}, we will analyze our system in terms of a phase space distribution 
$f(x^\mu, P_\mu)$ of fermions with positions $x^\mu$ and conjugate momenta $P_\mu$, and a classical 
scalar field $\phi(x^\mu)$. In the classical limit, all quantum operators $\hat{O}$ can be replaced by their
expectation values
\begin{equation}
    \langle \hat{O} \rangle \equiv \sum_s \int \frac{\mathrm{d}P_1
      \mathrm{d}P_2 \mathrm{d}P_3}{\sqrt{-\mathcal{G}}} \frac{1}{2
      P^0} f(x^\mu, P_\mu, s) \langle \phi, P^s | \hat{O} | \phi, P^s\rangle \, , \label{eq:expectation}
\end{equation}
where
\begin{itemize}
\item $P_i = \tilde{m} \frac{\mathrm{d}x_i}{\mathrm{d} \lambda}$ is the 
conjugate momentum  to the position $x^i$ of the fermions, with $\tilde{m} \equiv \sqrt{- P_\mu P^\mu}$ 
their mass and $\lambda$ their proper time.
\item $\mathcal{G}$ is the metric determinant.
\item $s$ is the fermion spin.
\item $\ket{\phi, P^s} \equiv \ket{\phi} \otimes \ket{P^s}$, with $\ket{\phi}$ a state with classical 
scalar field $\phi$ and $\ket{P^s}$ a one-particle fermion state with momentum $P$ and spin $s$. The 
former can be described by a coherent state~\cite{Sudarshan:1963ts,Glauber:1963tx,Glauber:1963fi}
\begin{equation}
    \ket{\phi} \equiv {\scalebox{1.5}{e}}^{\displaystyle -\frac{1}{2} \int \frac{\mathrm{d}^3 K}{(2 \pi)^3} \frac{|\phi(K)|^2}{(2K^0)^5}} {\scalebox{1.5}{e}}^{\displaystyle \,\int \frac{\mathrm{d}^3 K}{(2 \pi)^3} \frac{\phi(K)}{(2K^0)^{5/2}} {a_K^\phi}^\dagger } \ket{0} \, ,
\end{equation}
where
\begin{itemize}
\item $\phi(K)$ is the Fourier transform of the classical scalar field 
$\phi(x)$, i.e., 
\begin{equation}
\phi(x) \equiv \int \frac{\mathrm{d}^3K}{(2\pi)^3} \frac{1}{(\sqrt{2} K^0)^3} \left[\phi(K) e^{-i K x} + \phi(K)^* e^{iKx}\right] \, .
\end{equation}
\item $a_K^\phi$ is an annihilation operator of the field $\hat{\phi}$ 
with momentum $K$.
\item $\ket{0}$ is the vacuum.
\end{itemize}
The fermion one-particle state is given by~\cite{Peskin:1995ev}
\begin{equation}
    \ket{P^s} \equiv \sqrt{2 P^0} {a_P^s}^\dagger \ket{0} \, ,
\end{equation}
where $a_P^s$ is an annihilation operator of the field $\psi$ with 
momentum $P$.
\end{itemize}

We first start with the classical limit of \cref{eq:diracEq}. For convenience, we 
Fourier-expand the fermion field $\psi$ in terms of creation and 
annihilation operators following the conventions of
Ref.~\cite{Peskin:1995ev}, and \cref{eq:diracEq} reads\footnote{We are
also implicitly assuming that $\phi(x^\mu)$ can be considered to be
constant inside the coherence length of the fermion field $\Psi$.}
\begin{align}
a_P^s \left[\gamma_\mu P^\mu u^s(P) + (m_0 + g \hat{\phi}) u^s(P)\right] & = 0 \, , \label{eq:diracSpinor1}\\
b_P^{s \dagger} \left[\gamma_\mu P^\mu v^s(P) - (m_0 + g \hat{\phi}) v^s(P) \right] & = 0 \, , \label{eq:diracSpinor2}
\end{align}
with $b_P^s$ an antifermion annihilation operator and $\{u^s, v^s\}$ the  
spinor solutions to the Dirac equation. If we multiply
\cref{eq:diracSpinor1} by $a_P^{s \dagger} \left[u^{s \dagger}(P)
  \gamma_\mu P^\mu  - (m_0 + g \hat{\phi}) u^{s \dagger}(P)\right]$, and the 
Hermitian conjugate of \cref{eq:diracSpinor2} by $b_P^{s \dagger} \left[\gamma_\mu^\dagger P^\mu v^{s}(P) + (m_0 + g \hat{\phi}) v^{s \dagger}(P)\right]$, we get
\begin{align}
a_P^{s \dagger} a_P^s \, u^{s \dagger}(P) u^s(P) \left[-\tilde{m}^2 + (m_0 + g \hat{\phi})^2 \right] & = 0 \, ,\\
b_P^{s \dagger} b_P^{s} \, v^{s \dagger}(P) v^s(P) \left[-\tilde{m}^2 + (m_0 + g \hat{\phi})^2 \right] & = 0 \, .
\end{align}
We can now take classical expectation values using \cref{eq:expectation}. Since 
$\braket{\phi|\phi} = 1$ and $\braket{\phi|\hat{\phi}|\phi} = \phi$, the 
Dirac equations simply read
\begin{equation}
-\tilde{m}^2 + (m_0 + g \phi)^2 = 0 \, ,
\end{equation}
and so we have obtained the effective fermion mass
\begin{equation}
\tilde{m} = m_0 + g \phi \, .
\end{equation}

Finally, we can also take the expectation value of the scalar field 
equation~\eqref{eq:scalarMasterEq},
\begin{equation}
-D_\mu D^\mu \phi + M^2 \phi = - g \sum_s \int \frac{\mathrm{d}P_1
  \mathrm{d}P_2 \mathrm{d}P_3}{\sqrt{-\mathcal{G}}} \frac{1}{2 P^0}
f(x^\mu, P_\mu, s) \langle P^s | \bar{\psi} \psi| P^s\rangle \, .
\end{equation}
The expectation value on the right-hand side can be immediately evaluated
\begin{equation}
\langle P^s | \bar{\psi} \psi| P^s\rangle = \bar{u}^s(P) u^s(P) = 2 \tilde{m} \, .
\end{equation}
The same final result would be obtained if our state also contained antifermions. Thus, our final 
equation for the scalar field reads
\begin{equation}
-D_\mu D^\mu \phi + M^2 \phi = - g \int \frac{\mathrm{d}P_1 \mathrm{d}P_2 \mathrm{d}P_3}{\sqrt{-\mathcal{G}}} \frac{\tilde{m}}{P^0} f(x^\mu, P_\mu) \, ,
\end{equation}
where fermions, antifermions, and all spin orientations equally contribute to $f$.

To obtain the energy density and pressure of the system, we can
compute the expectation value of the stress-energy tensor using
\cref{eq:Lagrangian,eq:expectation}. The homogeneous and isotropic
results correspond to \cref{eq:rhobkg,eq:pbkg}.

\section{Properties of the Adiabatic Instability}
\label{sec:app_instability}

As discussed in \cref{sec:perturbations}, non-relativistic fermion density perturbations can grow exponentially 
under the presence of long range scalar interactions. In this
Appendix, we will approximately compute the fermion temperatures and 
interaction strengths for which this instability is present. We will also estimate the timescale over which non-linear 
\emph{nugget} formation takes place, and the conditions under which this happens much faster than cosmological scales. We will mostly follow the methodology in Ref.~\cite{Bjaelde:2007ki}.

From now on, we will assume that the adiabatic approximation~\eqref{eq:perturbedAdiabatic} always holds. As discussed in 
\cref{sec:perturbations}, this means that the inverse scalar effective mass ${\left[(k/a)^2 + M_\phi^2 + M_T^2\right]^{-1/2}}$ is much smaller than other timescales in the perturbed Klein Gordon equation~\eqref{eq:perturbedKG}. These timescales are
\begin{itemize}
    \item The Hubble scale, $H^{-1}$, which controls both the Hubble friction term as well as the timescale over which 
    the background quantities $\tilde{m}$ and $\varepsilon$ change.
    \item The timescale over which $\Psi_0$ changes. We will later check that this scale is $\gtrsim k/a$.
\end{itemize}
Since we are already assuming $M_\phi^2 + M_T^2 \gg H^2$ (see \cref{sec:bkgequations}), the adiabatic approximation holds
as long as $M^2 + M_T^2 \gg (k/a)^2$. In other words, we will solve the perturbation equations for physical length scales 
$a/k$ much larger than the interaction range. In addition, we will neglect metric perturbations in the Boltzmann 
equation~\eqref{eq:BoltzmannPerturb}. Using
\cref{eq:perturbedAdiabatic}, this equation then reads
\begin{equation}
    \frac{\partial \Psi(\vec{q}, \vec{k}, \tau)}{\partial \tau} + i \frac{\vec{k} \cdot \vec{q}}{\varepsilon(\tau)} \Psi(\vec{q}, \vec{k}, \tau) + i \frac{\vec{k} \cdot \vec{q}}{\varepsilon(\tau)} \frac{\tilde{m}(\tau)}{q^2} \frac{\mathrm{d} \log f_0}{\mathrm{d}\log q} \frac{g^2 \int \mathrm{d}^3q \, \frac{\tilde{m}(\tau)}{\varepsilon(\tau)} f_0(q) \Psi(\vec{q}, \vec{k}, \tau)}{(k/a)^2 + M_\phi^2 + M_T(\tau)^2} = 0 \, ,
    \label{eq:boltzmannEqAdiabatic}
\end{equation}
where $\vec{q} \equiv q \hat{n}$. This first order integro-differential equation cannot be solved in general. If, 
however, we consider timescales that are short with respect to cosmological evolution, the functions $\tilde{m}(\tau)$, 
$\varepsilon(\tau)$ and $M_T(\tau)$ can be assumed to be constant. We can then Fourier-transform in time
\begin{equation}
    \Psi(\vec{q}, \vec{k}, \tau) = \int \mathrm{d}\omega \, \tilde{\Psi}(\vec{q}, \vec{k}, \omega) e^{-i \omega \tau} \, ,
\end{equation}
and the equation reads
\begin{equation}
    - \omega \tilde{\Psi}(\vec{q}, \vec{k}, \omega) + \frac{\vec{k} \cdot \vec{q}}{\varepsilon} \tilde{\Psi}(\vec{q}, \vec{k}, \omega) + \frac{\vec{k} \cdot \vec{q}}{\varepsilon} \frac{\tilde{m}}{q^2} \frac{\mathrm{d} \log f_0}{\mathrm{d}\log q} \frac{g^2 \int \mathrm{d}^3q \, \frac{\tilde{m}}{\varepsilon} f_0 (q) \tilde{\Psi}(\vec{q}, \vec{k}, \omega)}{(k/a)^2 + M_\phi^2 + M_T^2} = 0\, ,
\end{equation}
or, rearranging terms,
\begin{equation}
    \tilde{\Psi}(\vec{q}, \vec{k}, \omega) = \left[ \int \mathrm{d}^3q \, \frac{\tilde{m}^2}{\varepsilon} f_0 (q) \tilde{\Psi}(\vec{q}, \vec{k}, \omega) \right] \frac{-g^2 \frac{\vec{k} \cdot \vec{q}}{q^2\varepsilon}\frac{\mathrm{d} \log f_0}{\mathrm{d}\log q}}{\left(-\omega + \frac{\vec{k} \cdot \vec{q}}{\varepsilon}\right)\left( k^2/a^2 + M_\phi^2 + M_T^2\right)} \, .
\end{equation}
To remove the dependence on $\tilde{\Psi}$, we multiply both sides by $\frac{\tilde{m}^2}{\varepsilon} f_0(q)$ and 
integrate over $\vec{q}$. Writing $\mathrm{d}^3q = 2\pi q^2 \mathrm{d}q \mathrm{d}(\vec{q} \cdot \vec{k})/k$, we can perform 
the angular integral and then integrate by parts, getting in the end
\begin{equation}
    \frac{(k/a)^2 + M_\phi^2 + M_T^2}{g^2 \tilde{m}^2} = 4 \pi \int_0^\infty \mathrm{d}q \frac{q^2 \tilde{m}^2}{\varepsilon^3\left[q^2 - \frac{\omega^2}{k^2} \varepsilon^2\right]} f_0(q) \, ,
    \label{eq:omegaperturb}
\end{equation}
an equation that gives $\omega = \omega(k, \frac{g m_0}{M_\phi}, T/M_0)$. Linear perturbations will be unstable if 
(and only if) this equation admits solutions with imaginary $\omega$. 

We will first study the existence of unstable solutions. As, for imaginary $\omega$, the right-hand side of 
\cref{eq:omegaperturb} is a monotonically growing function of $\omega^2$, there will be no unstable solutions if the 
left-hand side is greater than the right-hand side evaluated at $\omega^2 = 0$. That is, the system is unstable if and 
only if
\begin{equation}
    \frac{(k/a)^2 + M_\phi^2 + M_T^2}{g^2 \tilde{m}^2} \leq 4 \pi \int_0^\infty \mathrm{d}q \frac{\tilde{m}^2}{\varepsilon^3} f_0(q) \, .
    \label{eq:unstableCondition}
\end{equation}
This equation is quite instructive to understand the differences between this instability and the familiar Jeans 
gravitational instability. First, for ultrarelativistic fermions $\varepsilon \gg \tilde{m}$, the right-hand side will be supressed, and the system will 
generically be stable. That is, relativistic random thermal motions
stabilize perturbations at all scales for scalar
self interactions. This is different to the case of gravity, which has an 
infinite range (corresponding to $M_\phi^2 + M_T^2 \rightarrow 0$ in
\cref{eq:unstableCondition}) and thus for low enough $k$ there is always a scale, the 
Jeans scale, above which the accumulated gravitational attraction overcomes random thermal motions and 
perturbations collapse. Turning back to the scalar self interaction, even in the 
non-relativistic limit, for low enough fermion number densities the
right-hand side of \cref{eq:unstableCondition} decreases and the system is again 
stable. Physically, for interparticle distances larger than the interaction range scalar interactions turn off. Again, this 
is not the case for an infinite-range interaction as gravity, where
the left-hand side of \cref{eq:unstableCondition} can be made arbitrarily small by 
considering arbitrarily large scales. 

In order to obtain the 
temperatures and interaction strengths at which the system is
unstable, we have numerically solved \cref{eq:unstableCondition} in the limit $M_\phi^2 + M_T^2 \gg (k/a)^2$ (as discussed at the 
beginning of this Appendix) for the Fermi-Dirac fermion distribution
function in \cref{eq:bkgFD}. Our results are in shaded in
\cref{fig:instability}, where the blank region for which the system is stable corresponds to the two physical scenarios discussed above.

\begin{figure}[hbtp]
    \centering
    \includegraphics[width=0.85\textwidth]{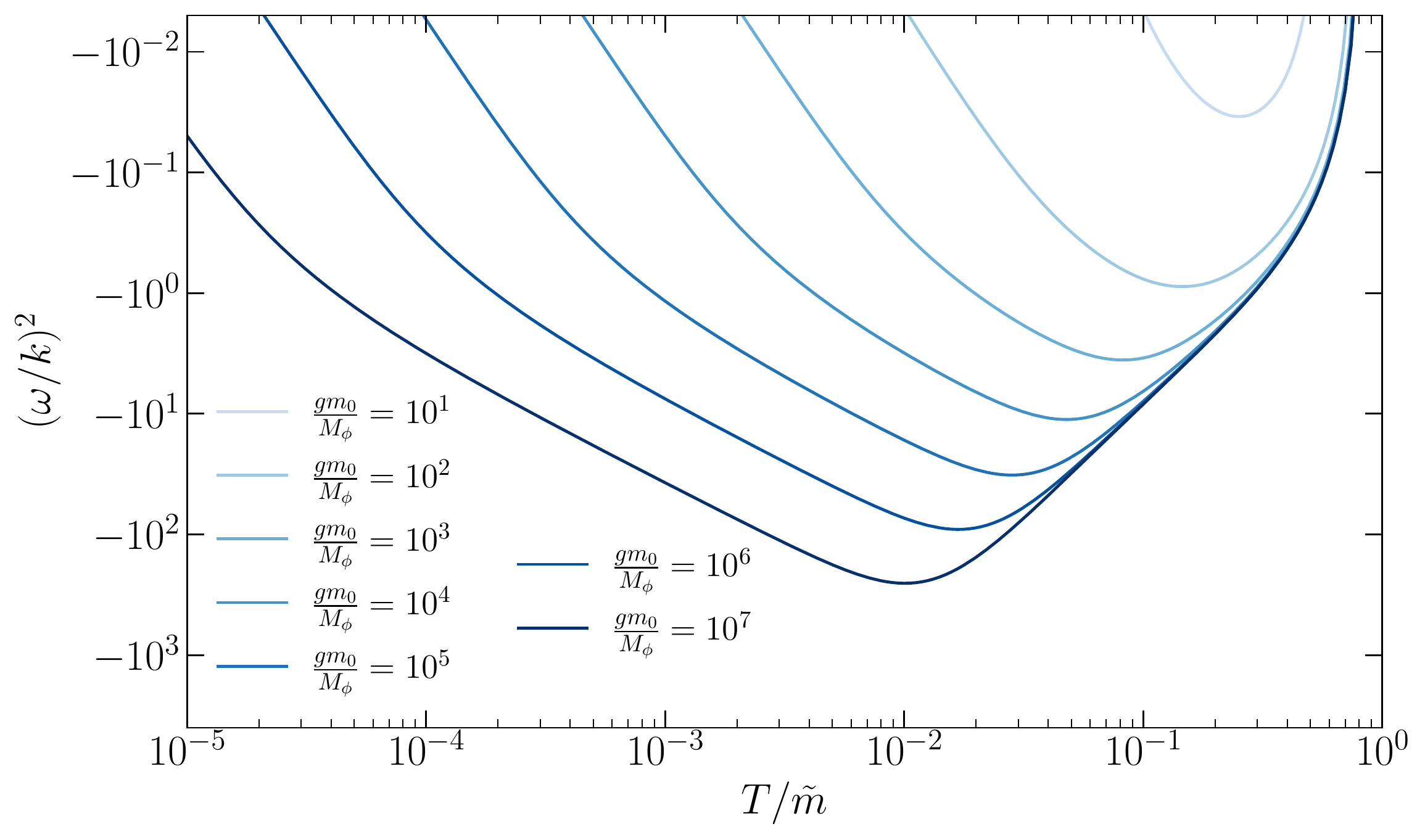}
    \caption{Perturbation growth rate as measured by $(\omega/k)^2$ as a function of $T/\tilde{m}$ for different interaction strengths [solid blue shades]. The curves cross $\omega/k = 0$ at the temperature of instability onset.}
    \label{fig:cssq}
\end{figure}

Apart from computing the temperatures and interaction strengths for which the system is unstable, \cref{eq:omegaperturb} 
also allows to estimate the timescale $\omega$ over which fermion density perturbations become non-linear and collapse in 
\emph{nuggets} as discussed in \cref{sec:perturbations}. To this purpose, we have numerically solved 
\cref{eq:omegaperturb} to obtain $(\omega/k)^2$ in the limit $M^2 + M_T^2 \gg (k/a)^2$ (as discussed at the 
beginning of this Appendix) and for the Fermi-Dirac fermion distribution function in \cref{eq:bkgFD}. We show in 
\cref{fig:cssq} $(\omega/k)^2$ as a function of the fermion temperature (normalized to its effective mass) for different 
interaction strengths $\frac{g m_0}{M_\phi}$. As we see, as soon as the system is unstable, $|\omega/k|$ quickly 
becomes $\mathcal{O}(1)$.

Nevertheless, under the assumptions in this Appendix, we can only compute perturbation growth for small timescales $\delta \tau$ 
during which all functions in \cref{eq:boltzmannEqAdiabatic} except for $\Psi$ can be considered to be constant. Over 
these timescales, unstable perturbations grow by a factor $e^{|\omega| \delta \tau}$. Since the initial perturbations are $\sim 10^{-4} \textup{--} 10^{-5}$, we will consider that fermion \emph{nuggets} form if
\begin{equation}
    |\omega| \delta \tau > \log(10^{4}\textup{--}10^5) \gtrsim 10 \, .
    \label{eq:nuggetConditionTau}
\end{equation}
In this case, \emph{nuggets} will form instantly with respect to cosmological time, and the instability can be modeled as 
an instantaneous transition to a dust-like behavior. A plausible value for $\delta \tau$ can be estimated as follows: if 
$\omega(\tau)$ as computed from \cref{eq:omegaperturb} is constant, then all relevant parameters in that equation are 
also constant. We have thus taken $\delta \tau$ as a fraction $\varepsilon$ of the typical time during which 
$\omega(\tau)$ changes
\begin{equation}
    \delta \tau = \varepsilon \left| \frac{\omega}{\mathrm{d}\omega/\mathrm{d}\tau}\right| \, ,
\end{equation}
where $\mathrm{d}\omega/\mathrm{d}\tau$ can be obtained from \cref{fig:cssq}. By conservatively assuming $\varepsilon = 
10^{-2}$ and $k/a = 0.1 \sqrt{M_\phi^2 + M_T^2}$, our condition~\eqref{eq:nuggetConditionTau} for instantaneous nugget 
formation is equivalent to
\begin{equation}
    \sqrt{M_\phi^2 + M_T^2} \gtrsim 10^5 H \, ,
\end{equation}
at instability onset. This corresponds to \cref{eq:nugget_condition} in the main text.

\section{Statistical Analysis in the Whole Parameter Space}
\label{sec:app_triangles}

\begin{figure}[hbtp]
    \centering
    \includegraphics[width=\textwidth]{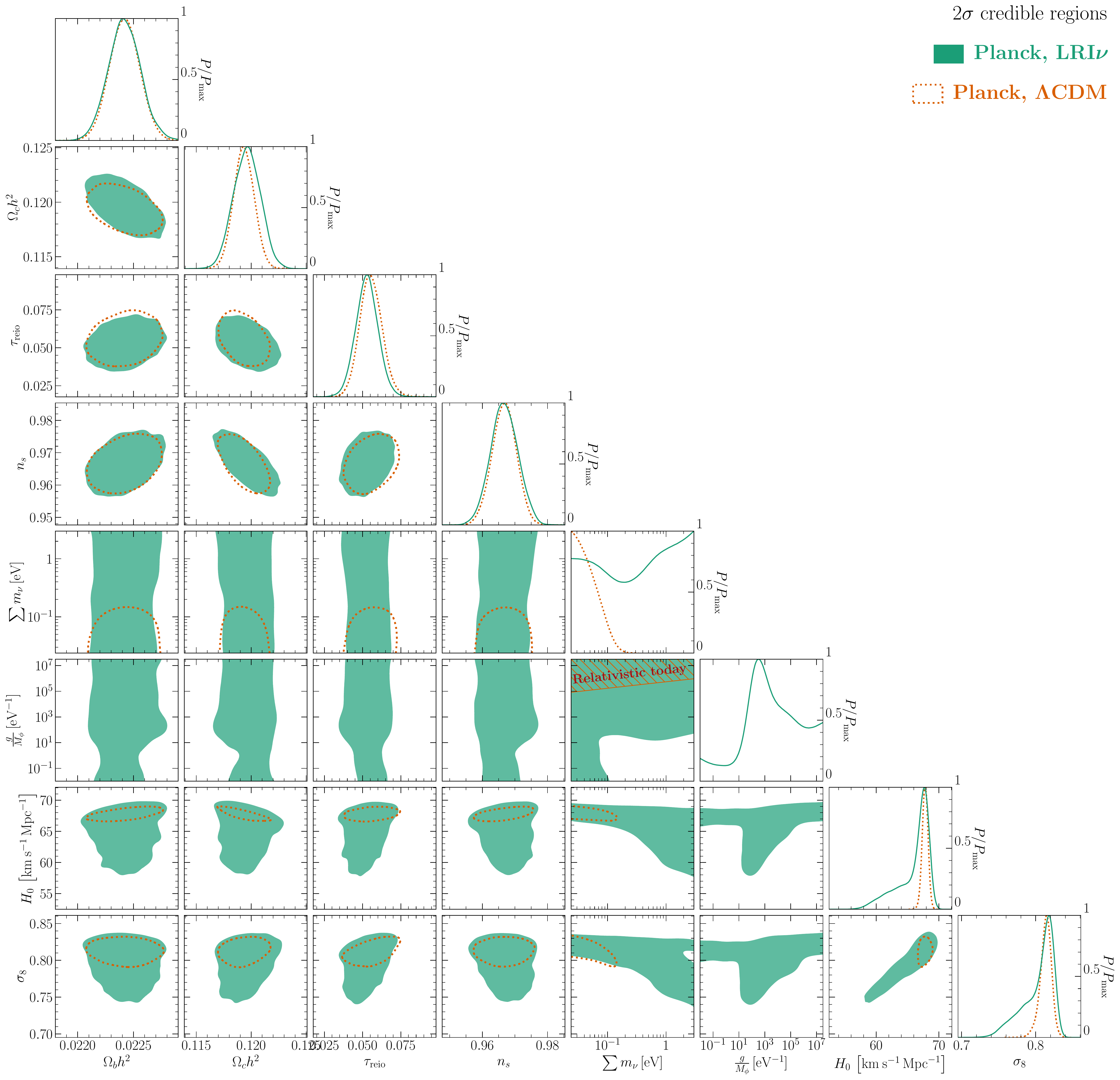}
    \caption{Planck 2018 constraints~\cite{Aghanim:2018eyx} on neutrino long range interactions [solid light green] and on $\Lambda$CDM with non-interacting massive neutrinos [dotted orange]. We show the marginalized $2\sigma$ credible regions and 1-D posterior probability distributions. For $\sum m_\nu$ and $\frac{g}{M_\phi}$, posteriors are constructed with uniform logarithmic bins. In the hatched region, cosmic neutrinos are relativistic today. The dark green line is the minimum value of $\sum m_\nu$ allowed by neutrino oscillation data~\cite{Esteban:2020cvm,deSalas:2020pgw,Capozzi:2017ipn}.}
    \label{fig:triangle_planck_full}
\end{figure}

\begin{figure}[hbtp]
    \centering
    \includegraphics[width=\textwidth]{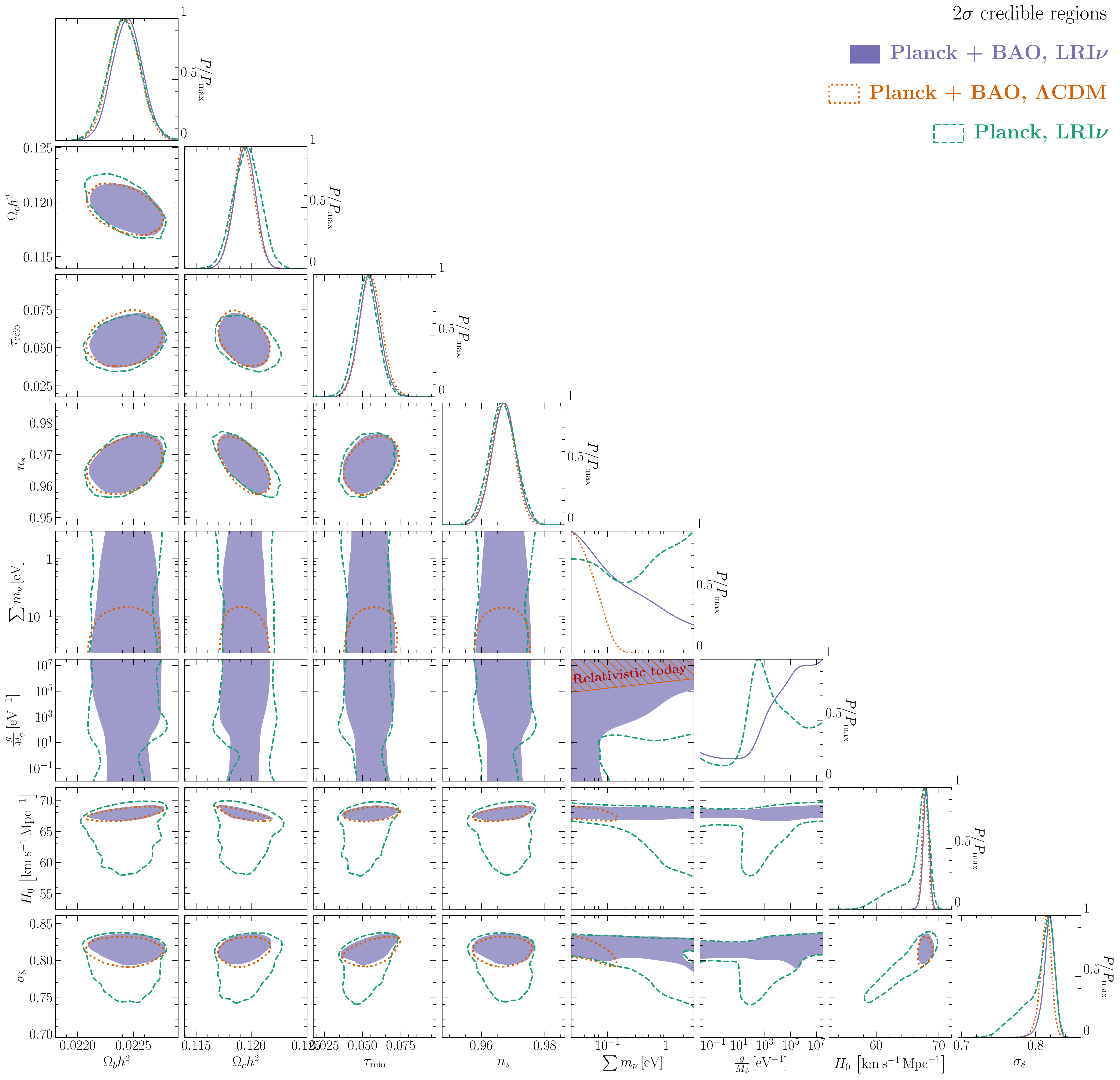}
    \caption{
    Planck 2018+BAO constraints~\cite{Aghanim:2018eyx,Beutler:2011hx,Ross:2014qpa,Alam:2016hwk} on 
    neutrino long range interactions [solid purple], on $\Lambda$CDM with non-interacting massive neutrinos 
    [dotted orange], and Planck 2018 constraints~\cite{Aghanim:2018eyx} on neutrino long range interactions 
    [dashed green]. We show the marginalized $2\sigma$ credible regions and 1-D posterior probability 
    distributions. For $\sum m_\nu$ and $\frac{g}{M_\phi}$, posteriors are 
    constructed with uniform logarithmic bins. In the hatched region, cosmic neutrinos are relativistic 
    today. The solid dark green line is the minimum value of $\sum m_\nu$ allowed by neutrino oscillation data~\cite{Esteban:2020cvm,deSalas:2020pgw,Capozzi:2017ipn}.}
    \label{fig:triangle_planck_BAO_full}
\end{figure}

\begin{figure}[hbtp]
    \centering
    \includegraphics[width=\textwidth]{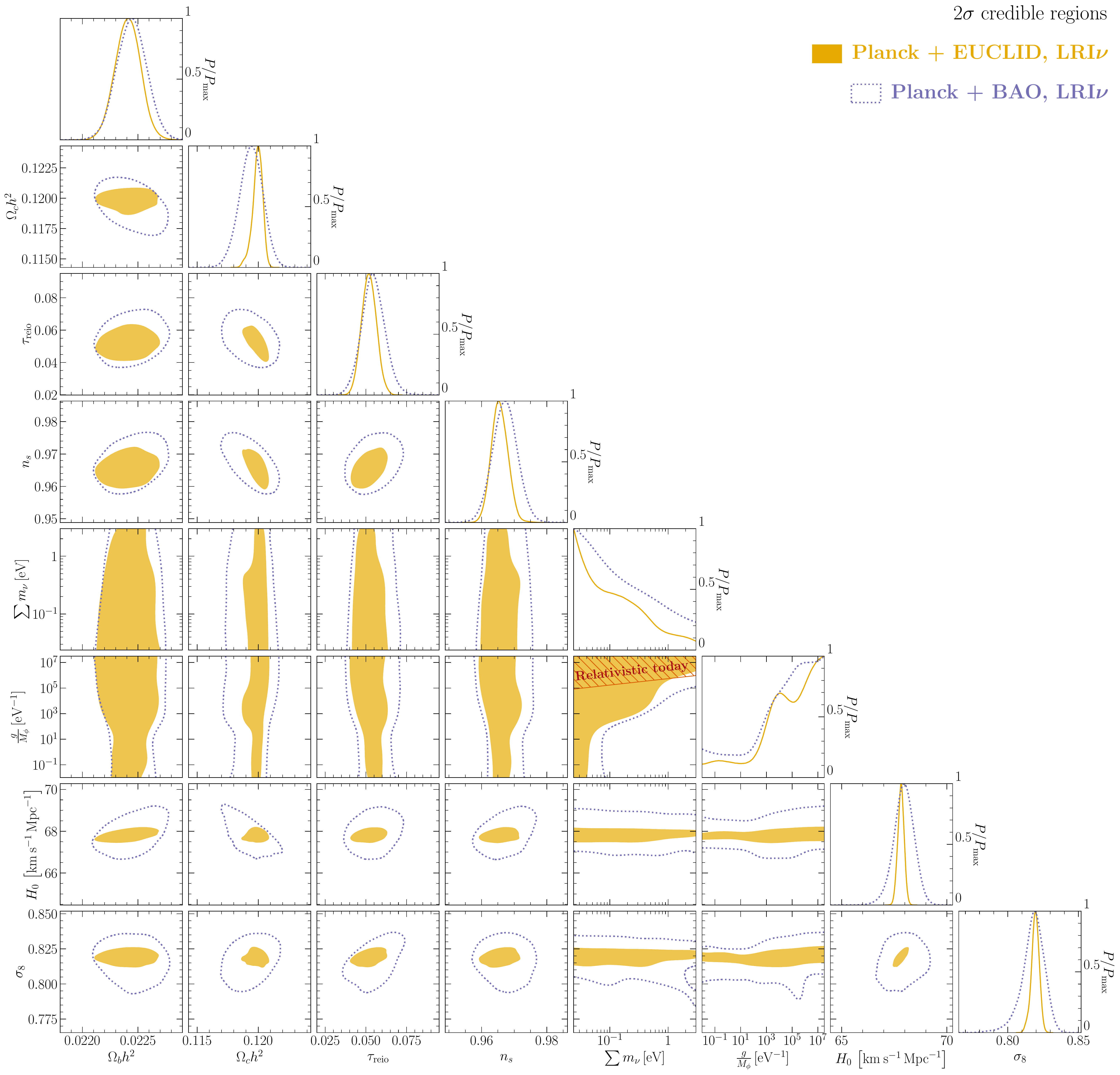}
        \caption{Future EUCLID + Planck 2018 constraints~\cite{Laureijs:2011gra,Sprenger:2018tdb,Aghanim:2018eyx} on 
      long range interacting neutrinos, LRI$\nu$, [solid yellow], and Planck 2018 + BAO 
    constraints~\cite{Aghanim:2018eyx,Beutler:2011hx,Ross:2014qpa,Alam:2016hwk} [dotted purple]. We show the 
    marginalized $2\sigma$ credible regions and 1-D posterior probability distributions. For $\sum m_\nu$ and $\frac{g}{M_\phi}$, posteriors are constructed with uniform logarithmic bins. In the hatched region, cosmic 
    neutrinos are relativistic today. We have generated EUCLID mock data with the best fit parameters from the 
    $\Lambda$CDM Planck 2018 + BAO analysis (last column in Table 2 in Ref.~\cite{Aghanim:2018eyx}), no long range 
    interactions and $\sum m_\nu = $ 0.024 eV, 
    the smallest value compatible with our priors, and consistent with massless neutrinos within EUCLID precision.}
    \label{fig:triangle_euclid_full}
\end{figure}

\begin{figure}[hbtp]
    \centering
    \includegraphics[width=\textwidth]{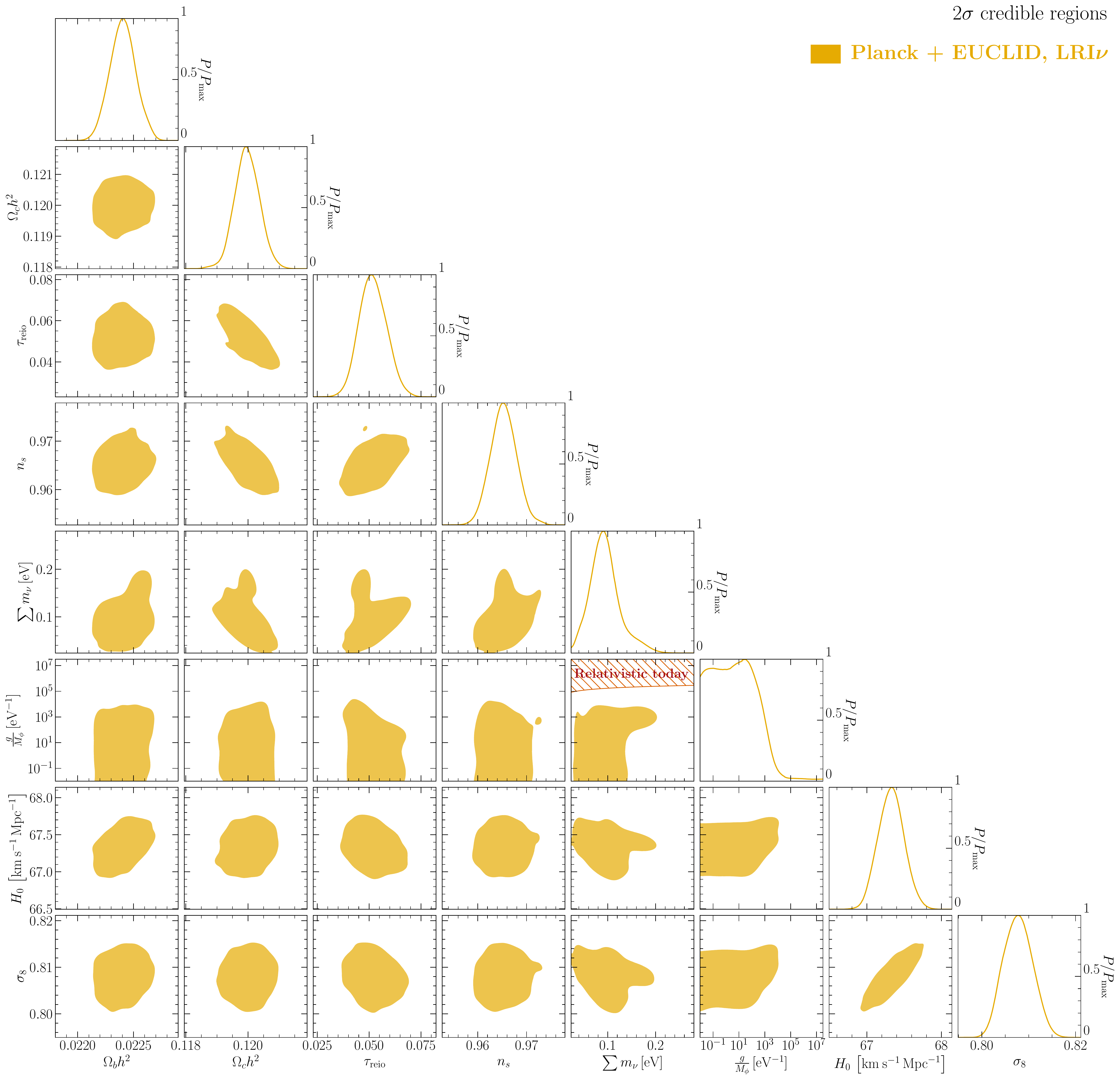}
        \caption{Future EUCLID + Planck 2018 constraints~\cite{Laureijs:2011gra,Sprenger:2018tdb,Aghanim:2018eyx} on 
      long range interacting neutrinos, LRI$\nu$, [solid yellow]. We show the 
    marginalized $2\sigma$ credible regions and 1-D posterior probability distributions. For $\frac{g}{M_\phi}$, posteriors are constructed with uniform logarithmic bins. The prior on $\sum m_\nu$ is taken to be linear. In the hatched region, cosmic 
    neutrinos are relativistic today. We have generated EUCLID mock data with the best fit parameters from the 
    $\Lambda$CDM Planck 2018 + BAO analysis (last column in Table 2 in Ref.~\cite{Aghanim:2018eyx}), no long range 
    interactions and $\sum m_\nu = $ 0.08 eV, 
    a value compatible with CMB and BAO data that should be well-measurable by EUCLID.}
    \label{fig:triangle_euclid_full_massive}
\end{figure}

In this Appendix, we show the results of our Bayesian analysis 
(\cref{fig:triangle_planck,fig:triangle_planck_BAO,fig:EuclidTriangle}) for all cosmological parameters. 
\Cref{fig:triangle_planck_full} corresponds to the analysis of Planck 2018 data, \cref{fig:triangle_planck_BAO_full} also 
includes BAO data, and \cref{fig:triangle_euclid_full,fig:triangle_euclid_full_massive} corresponds to the analysis prospects of EUCLID and Planck 2018 
data. See the main text for the description of the analysis.

\clearpage
\bibliographystyle{JHEP}
\bibliography{longrange}
\end{document}